\documentclass{emulateapj}

\usepackage{epsfig,amsmath,amssymb,amsfonts,natbib,color,url}  

\usepackage{amsmath,amssymb,amsfonts,natbib}
\usepackage {graphicx,color}
\usepackage {txfonts }
\usepackage {natbib  }
\usepackage {float   }
\usepackage {enumerate}
\usepackage {subfigure  }
\providecommand{\e}[1]{\ensuremath{\times 10^{#1}}}
\newcommand{\sub}[1]{_{\rm{#1}}}

\newcommand{\trad}{\ensuremath{\tau\sub{rad}}}
\newcommand{\tdrag}{\ensuremath{\tau\sub{drag}}}
\newcommand{\tadv}{\ensuremath{\tau\sub{adv}}}
\newcommand{\twave}{\ensuremath{\tau\sub{wave}}}
\newcommand{\tvert}{\ensuremath{\tau\sub{vert}}}
\newcommand{\forcing}{\ensuremath{\Delta h\sub{eq}/H} }

\newcommand{\revd}[1]{{#1}}

\def\tt{\bf }

\shorttitle{Heat redistribution on hot Jupiters}
\shortauthors{Perez-Becker \& Showman}

\begin{document}

\shorttitle{} \shortauthors{Perez-Becker \& Showman}

\title{ATMOSPHERIC HEAT REDISTRIBUTION ON HOT JUPITERS}

\author{Daniel Perez-Becker\altaffilmark{}} \affil{Department of Physics, University of California, Berkeley, CA 94720, USA}
\and

\author{Adam P. Showman\altaffilmark{}} \affil{Department of Planetary Sciences, Lunar and Planetary Laboratory, University of Arizona, Tucson, AZ 85721, USA}

\begin{abstract} \revd{Infrared light curves of transiting hot Jupiters  
present a trend in which the atmospheres of the hottest planets are less
efficient at redistributing the stellar energy absorbed on their
daysides---and thus have a larger day-night temperature contrast---than
colder planets. To this day, no predictive atmospheric model has been
published that identifies which dynamical mechanisms determine the atmospheric heat redistribution efficiency on tidally locked exoplanets. 
Here we present a shallow water model of
the atmospheric dynamics on synchronously rotating planets that
explains why heat redistribution efficiency drops as stellar insolation rises. 
Our model
shows that planets with weak friction and weak irradiation exhibit a
banded zonal flow with minimal day-night temperature differences,
while models with strong irradiation and/or strong friction exhibit a
day-night flow pattern with order-unity fractional day-night
temperature differences. 
To interpret the model, we develop a scaling theory which shows that the
timescale for gravity waves to propagate horizontally over planetary scales, $\twave$, plays a dominant role in controlling the transition from small to large temperature contrasts.
This implies that heat redistribution is governed by a wave-like process, similar to the one responsible for the weak temperature gradients in the Earth's tropics. When atmospheric drag can be neglected, the transition from small to large day-night temperature contrasts occurs when $\twave \sim \sqrt{\trad/\Omega}$, where $\trad$ is the radiative relaxation time and $\Omega$ is the planetary rotation frequency. 
Alternatively, this transition criterion can be expressed as $\trad \sim \tvert$, where $\tvert$ is the timescale for a fluid parcel to move vertically over the difference in day-night thickness. These results subsume the more widely used timescale comparison for estimating heat redistribution efficiency between $\trad$ and the horizontal day-night advection timescale, $\tadv$. 
Only because $\tadv \sim \tvert$ for hot Jupiters does the commonly assumed timescale comparison between $\trad$ and $\tadv$ yield approximately correct predictions for the heat redistribution efficiency.} 
\end{abstract}

\section{INTRODUCTION} 
\label{sec-intro}

Stellar radial velocity surveys have discovered a class of extrasolar
planets whose masses are comparable to Jupiter's, but that orbit their host
stars at distances less than 0.1 AU \citep{Lovis:2010p17319}. In
contrast to Jupiter, which roughly receives as much power from stellar
irradiation as it releases from its interior, these ``hot Jupiters''
have power budgets dominated by external irradiation. In addition,
the strong tidal torques are presumed to lock them into a state
of synchronous rotation \citep{Guillot:1996p17309}, forcing them to
have permanent daysides and nightsides.  The extreme insolation, fixed
day-night thermal forcing pattern, and slow rotation rates of hot
Jupiters provide a unique laboratory to explore the atmospheric
dynamics of giant planets under conditions not present in the Solar
System.

Continuous photometric observations of eclipsing (transiting) hot
Jupiters, over half an orbit or longer, allow for precise
determination of the flux emitted from the planet as it presents
different phases to us. Emission from the nightside is visible around
the time the planet transits the stellar disk, while the dayside is
visible just before and after the planet passes behind the star
(secondary eclipse). The planetary contribution to the total flux
received is up to $\sim$0.1\%--0.3\% at infrared wavelengths and can be
isolated because the precise stellar flux is known from observations
during secondary eclipse. The longitudinal atmospheric temperature
profile of the planet is inferred from the orbital flux
variations. These light curve observations are currently one of the
most powerful tools to constrain the atmospheric dynamics of these
planets \citep[for a review, see][]{Deming:2009p17560}.
So far, visible and infrared light curves for at least 11 hot
Jupiters have been published.

\begin{figure}
\includegraphics[width=\linewidth]{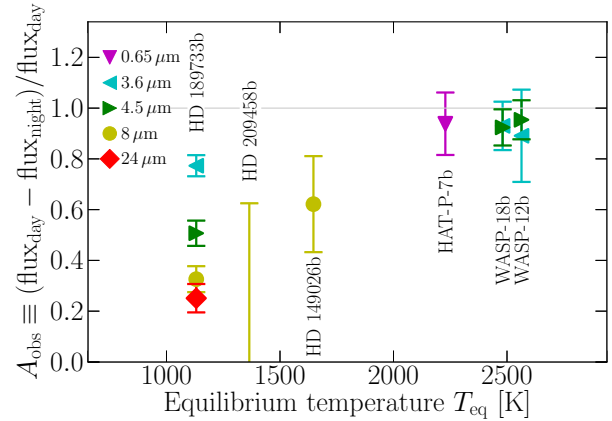}
\caption{\revd{Fractional day-night infrared flux variations $A\sub{obs}$
  vs. global-mean equilibrium temperature $T\sub{eq}$ for hot
  Jupiters with measured light curves.  Here, $A\sub{obs}$ is defined,
  at a particular wavelength, as the flux difference between dayside
  and nightside divided by the dayside flux.  The equilibrium
  temperature is defined as $T\sub{eq} = [F_*/(4\sigma)]^{1/4}$,
  where $F_*$ is the stellar flux received by the planet and
  $\sigma$ is the Stefan-Boltzmann constant. Planets with hotter mean
  temperatures have larger day-night flux variations, indicating
  larger longitudinal temperature gradients at the photosphere. Colored symbols
  with error bars are from published observations
  \citep{Knutson:2007p12609,
     Knutson:2009p17915, Knutson:2009p14854, Knutson:2012p14994, Cowan:2007p14836,Cowan:2012p14586, Borucki:2009p14771, Maxted:2013p17382}. The error bar without a data point for HD
  209458b indicates that only an upper limit was published
  \citep{Cowan:2007p14836}.}}
\label{fig-obs-nomodel}
\end{figure}

Figure~\ref{fig-obs-nomodel} presents the fractional day-night flux
differences, obtained from such light curves, for transiting hot
Jupiters on near-circular orbits.\footnote{In practice, we are
  dividing the amplitude of the phase curve variation by the \revd{depth of
the secondary eclipse relative to the maximum in the phase curve}, a procedure 
that only works for transiting hot
  Jupiters.  Light curves of non-transiting hot Jupiters such as Ups
  And b \citep{Harrington:2006p14796, Crossfield:2010p12798}, 51 Peg b,
  and HD 179949b \citep{Cowan:2007p14836} are therefore not included in
  Figure~\ref{fig-obs-nomodel}.  We also exclude planets on highly
  eccentric orbits such as HD 80606b \citep{Laughlin:2009p15306} and
  HAT-P-2b \citep{Lewis:2013p17424}.}  Specifically, we plot the flux
difference between the dayside and nightside, divided by the dayside
flux, as a function of the planet's global-mean equilibrium
temperature calculated assuming zero albedo.  Interestingly, these
light curve observations suggest an emerging trend wherein planets
that receive greater stellar insolation---and therefore have hotter
mean temperatures---exhibit greater fractional flux contrasts between
the dayside and the nightside.  Cool planets like HD 189733b
\citep{Knutson:2007p12609, Knutson:2009p14854, Knutson:2012p14994} and
HD 209458b \citep{Cowan:2007p14836} have only modest day-night flux
differences.  Planets receiving intermediate flux, such as HD 149026b
\citep{Knutson:2009p17915},
exhibit intermediate day-night flux differences.  And the most
strongly irradiated planets, such as HAT-P-7b
\citep{Borucki:2009p14771}, WASP-18b \citep{Maxted:2013p17382}, \revd{and
WASP-12b \citep{Cowan:2012p14586}}
exhibit fractional day-night flux differences close to
unity.\footnote{The HAT-P-7b point in Figure~\ref{fig-obs-nomodel}
  lies in the visible, so there remains ambiguity about whether the
  dayside is bright due to thermal emission or due to reflected
  starlight.  However, an unpublished full-orbit light curve of
  HAT-P-7b at $3.6\,\mu$m from Spitzer exhibits a phase-curve
  amplitude comparable to the secondary-eclipse depth \citep{Knutson:2011p17303},
  where emission is predominantly thermal rather than reflected
  starlight.  This demonstrates that HAT-P-7b indeed exhibits large
  fractional day-night temperature contrasts at the photosphere.}  This
trend is consistent with a comparison of secondary eclipse depths with
predicted dayside equilibrium temperatures in a broad sample of $\sim$24
systems performed by \citet{Cowan:2011p14321}.  Together,
Figure~\ref{fig-obs-nomodel} and the analysis of
\citet{Cowan:2011p14321} suggest that the day-night temperature
difference at the photosphere (expressed as a ratio to the dayside
temperature) increases with effective temperature, and approaches
unity for hot Jupiters with effective temperatures of
$\sim$$2200\rm\,K$ or greater.

What causes the trend observed in Figure \ref{fig-obs-nomodel}? The
commonly invoked explanation has been that the efficiency with which hot
Jupiters redistribute heat is determined by the extent to which
atmospheric winds transport hot gas across planetary scales faster
than it takes the gas to radiate its heat into space. This balance is
typically described by a comparison between two characteristic
timescales, the timescale for winds to advect gas horizontally 
across the planet,
$\tadv$, and the timescale for gas to reach local radiative
equilibrium, $\trad$. \citet{Showman:2002p12764} first suggested
that hot Jupiters will exhibit large fractional day-night temperature
differences when $\trad\ll \tadv$ and small fractional day-night
temperature differences when $\trad \gg \tadv$.  They pointed out that
the radiative time constant decreases strongly with increasing
temperature, and they presented a heuristic theory suggesting that
planets with greater characteristic radiative heating rates will
exhibit larger fractional day-night temperature
differences. Subsequently, a wide range of authors proposed that this
timescale comparison could describe the pressure-, opacity-, and
insolation-dependence of the day-night temperature differences on hot
Jupiters \citep{Cooper:2005p13046, Showman:2008p17307,
  Fortney:2008p17508, Rauscher:2010p14649, Heng:2011p17534,
  Cowan:2011p14321, Perna:2012p17536}.  This picture is based on the
expectation, from both three-dimensional (3D) circulation models and
observations, that hot Jupiters should develop fast atmospheric jets
capable of transporting heat over planetary scales.

However, this timescale comparison is neither self-consistent nor
predictive, as $\tadv$ is not known a priori and depends on many
atmospheric parameters, including $\trad$.  In particular, one cannot
even evaluate the criterion under given external forcing conditions
unless one already has a theory for (or simulations of) the
atmospheric flow itself.  No such theory for the atmospheric
circulation generally, or the day-night temperature difference
specifically, has ever been presented.  Moreover, the comparison
between $\tadv$ and $\trad$ neglects any role for other important
timescales, including timescales for wave propagation, frictional
drag, and planetary rotation. These timescales almost certainly
influence the day-night temperature difference, among other aspects
of the circulation.  More generally, it is crucial
to emphasize that the ultimate goal is not simply to obtain a 
timescale comparison for the transition between small and large day-night
temperature difference, but rather to obtain a predictive theory for the 
day-night temperature difference itself, valid across the full
parameter space.

To reiterate the role that other timescales can play, consider the
Earth's tropics.  Over most of the tropics, the horizontal temperature
gradients are weak, and the radiative cooling to space is balanced not
by horizontal advection but by {\it vertical} advection
\citep[e.g.,][]{Sobel:2001p14821}---raising questions about
the relevance of the horizontal advection time for this system.  In
fact, the longitudinal variation of the temperature structure in the tropics is
primarily regulated by adjustment due to gravity waves \citep[see][for
  a review]{Showman:2013p17828}.  Moist convection, gradients in
radiative heating, and other processes can lead to horizontal
temperature and pressure variations that in turn cause the generation
of gravity waves.  These waves induce horizontal
convergence/divergence that, via mass continuity, causes air columns
to stretch or contract vertically.  This coherent vertical motion
pushes isentropes up or down, and if the Coriolis force is weak (as it is in the
tropics) and the waves are able to radiate to infinity, the final state
is one with flat isentropes. A state with
flat isentropes is a state with constant temperature on isobars;
therefore, this wave-adjustment process acts to erase horizontal
temperature differences.
As emphasized by \citet{Bretherton:1989p15029} and others,
this adjustment process occurs on characteristic timescales comparable
to the time for gravity waves to propagate over the length scale of
interest. A key point is that, in many cases, this wave-propagation
timescale is much shorter than the timescales for horizontal advection
or mixing (e.g., between a cumulus cloud and the surrounding
environment).  Such wave propagation is not only local but 
can also occur over planetary scales, on both Earth
\citep[e.g.,][]{Matsuno:1966p0001, Gill:1980p12543, Bretherton:2003p15030} and
exoplanets \citep{Showman:2011p12973}.  Therefore, it is reasonable
to expect that this wave-adjustment process will play a key role in
the regulation of horizontal temperature differences on exoplanets---and,
as a corollary, that the horizontal wave-propagation timescale will be important.

Motivated by these issues, our goals are to (1) quantify in numerical
simulations how the day-night temperature difference on synchronously
rotating exoplanets depends on external forcing parameters, (2)
develop a quantitative theory for this behavior, and (3) isolate the
dynamical mechanisms responsible for controlling the day-night
temperature differences.  A natural spin-off of this undertaking
will be a criterion, expressed in terms of the various timescales, for the
transition between small and large day-night temperature differences.

Because our emphasis is on developing a basic understanding, we adopt
perhaps the simplest dynamical model that can capture the key physics:
a shallow-water model representing the flow in the observable
atmosphere. This means that our model will exclude many details
important on hot Jupiters, but it will allow us to identify the key
dynamical processes in the cleanest possible environment.  We view
this as a prerequisite to understanding more realistic systems.

Section~\ref{sec-model} introduces our dynamical model.
Section~\ref{sec-numerical-solutions} presents numerical solutions of
the atmospheric circulation and explores the dependence on external
forcing parameters.  Section~\ref{sec-theory} presents an analytic
scaling theory of the day-night differences explaining the behavior
of our numerical solutions.  Section~\ref{sec-discussion}
provides a dynamical interpretation of the behavior in our theory and
simulations. Section~\ref{sec-observations} applies this understanding
to observations, and in particular we show that our model can explain the
trend observed in Figure~\ref{fig-obs-nomodel}. Finally,
Section~\ref{sec-conclusions} concludes.

\section{THE MODEL}
\label{sec-model}

\begin{figure}
\includegraphics[width=\linewidth]{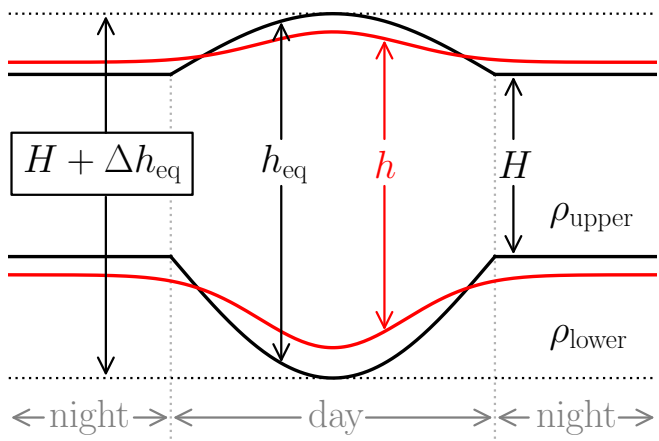}
\caption{%
The two-layer shallow water model. A layer of fluid with constant density $\rho\sub{upper}$ and variable thickness $h$ floats above an infinitely deep layer of fluid with higher constant density $\rho\sub{lower}$. The fluid thickness $h$ (red solid lines) represents the atmospheric mass column that has an entropy above a given reference value. Atmospheric heating thus will locally increase $h$, while cooling will locally reduce $h$. Radiative relaxation tends to restore the fluid to a temporally invariant profile, $h\sub{eq}$ (black solid lines), over a radiative relaxation timescale $\trad$. The radiative equilibrium profile, defined in equation (\ref{2Dheq}), varies in height from $H$ on the nightside to $H +\Delta h\sub{eq}$ at the substellar point.}
\label{fig-sw}
\end{figure}

Global, 3D circulation models (GCMs) of planetary atmospheres involve many interacting processes, which makes it difficult to identify which
dynamical mechanisms are dominating the solution. Simplified models
have therefore played an important role in atmospheric dynamics of
giant planets, both solar \citep{Dowling:1989p14864, Cho:1996p14865,
  Scott:2007p14868, Showman:2007p14871} and
extrasolar \citep{Cho:2003p17276, Cho:2008p17275, Showman:2011p12973, Showman:2013p17094}.

We study atmospheric heat transport on hot Jupiters with an idealized
two-layer shallow-water model (see Figure \ref{fig-sw}). The buoyant
upper layer of the model, having constant density $\rho\sub{upper}$
and variable thickness $h$, represents the meteorologically active
atmosphere of the planet. The infinitely deep bottom layer has a
higher constant density $\rho\sub{lower}$ and represents the
convective interior of the planet. We assume isostasy: the total mass
column above a given depth in the lower layer is constant.  For an
infinitely deep lower layer, isostasy is guaranteed for baroclinic
waves---those where the upper free surface and the interfacial
boundary bow in opposite directions (Figure \ref{fig-sw}).
Isostasy captures baroclinic modes but screens out barotropic ones
(see Section 6.2 of \citealt{Gill:1982p0001}); isostasy implies that there are no
horizontal pressure gradients in the lower layer and therefore no
horizontal velocities there. However, there can be vertical
velocities; indeed mass will be transferred between the two layers.

The equations governing the upper layer are

\begin{equation}\label{2DShallowWaterA}
\frac{d \bf{v}}{dt}+g \nabla h +f \bf{k}\times \bf{v}= \bf{R} - \frac{\bf{v}}{\tau_{\rm{drag}}},
\end{equation}
\begin{equation}\label{2DShallowWaterB}
\frac{\partial h}{\partial t}+ \nabla \cdot ( \textbf{v} h )= \frac{h_{\rm{eq}}(\lambda,\phi)-h}{\tau_{\rm{rad}}} \equiv Q,
\end{equation}

\noindent where $\textbf{v}(\lambda, \phi,t)$ is the horizontal
velocity, $h(\lambda,\phi,t)$ is the local thickness, $t$ is time, $g$
is the reduced gravity,\footnote{The reduced gravity is the local
  gravitational acceleration times the fractional density difference
  between the upper and lower layers, $g\equiv GM_{\rm{planet}}/a^2
  \times (\rho_{\rm{lower}}-\rho_{\rm{upper}})/\rho_{\rm{lower}}$,
  where $G$ is the gravitational constant, $M_{\rm planet}$
    is the planet mass, and $a$ is the planet radius. Note that our
  definition of $g$ differs from $g'$ in Vallis
  (\citeyear{Vallis:2006p12972}, see their Section 3.2),
  with $g=g'\times \rho\sub{lower}/\rho\sub{upper}$. The reduced
  gravity measures the restoring force at the interface between the
  two layers. When both layers have the same density, no pressure
  gradient forces exist and $g=0$. Similarly, when $\rho\sub{upper}
  \ll \rho\sub{lower}$, $g$ will equal the full gravitational
  acceleration.} $f=2 \Omega \sin{\phi}$ is the Coriolis parameter,
$\textbf{k}$ is the vertical unit vector, $\nabla$ is the
  horizontal gradient operator, $\Omega$ is the planetary rotation
frequency, and $(\lambda,\phi)$ are the longitudinal and latitudinal
angles. Here $d/dt \equiv \partial /\partial t + \textbf{v} \cdot
\nabla$ is the time derivative following the flow (this includes
curvature terms in spherical geometry).

\begin{figure*}
\includegraphics[width=7.0in]{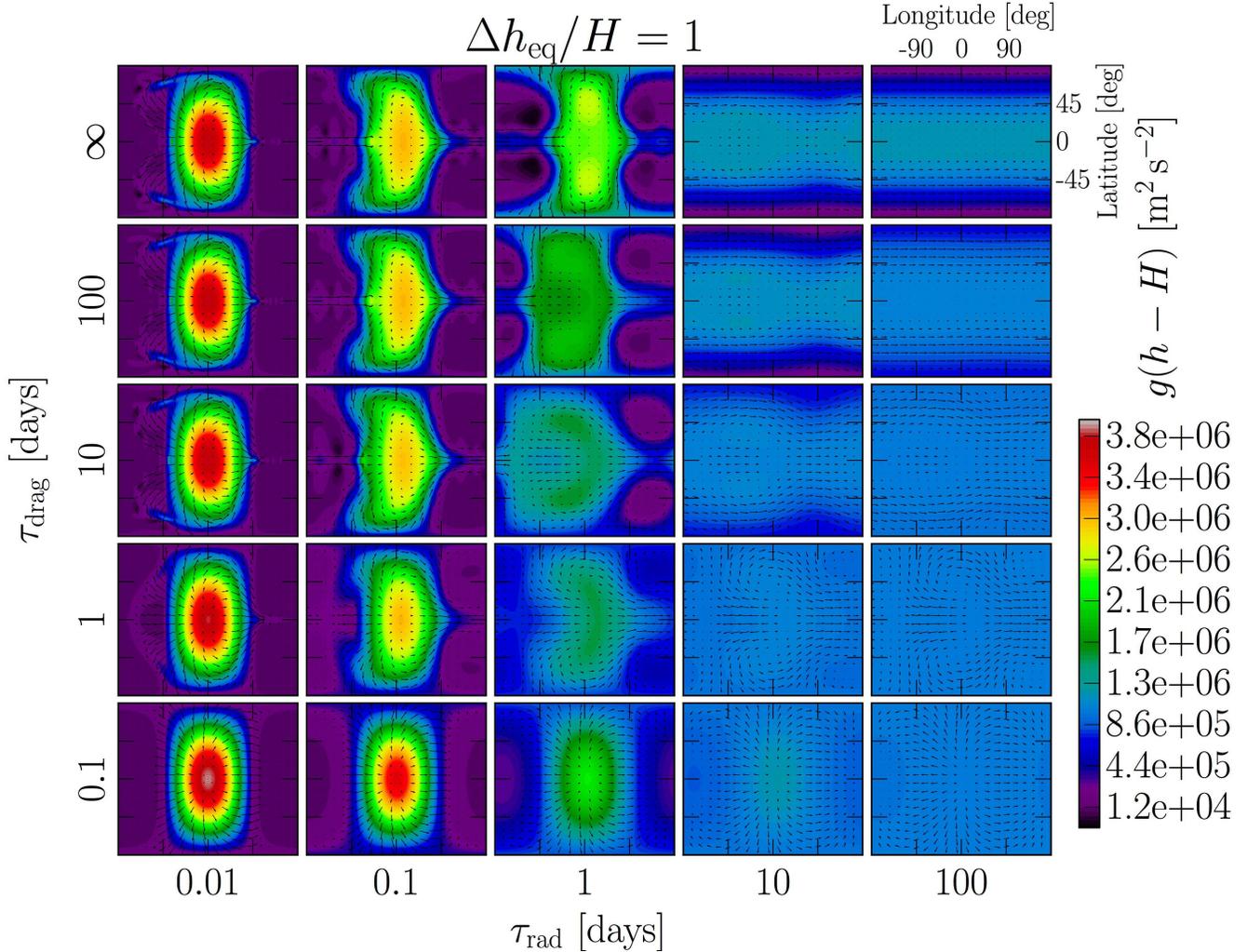}
\caption{%
Equirectangular maps of steady-state geopotential
  ($gh$) contours for the equilibrated solutions of the shallow-water
  model for a fractional forcing amplitude of $\Delta
  h\sub{eq}/H=1$. We have subtracted the constant value of $gH=4\times{10}^6$~m$^2$~s$^{-2}$ from each panel. Overplotted are vector fields of the steady-state
  winds. Each panel in the grid was computed for a different combination of
radiative and drag timescales, $\trad$ and $\tdrag$, expressed in Earth days. 
Panels share the same scale for the geopotential, but wind
  speeds are normalized independently in each panel. The substellar
  point is located at the center of each panel, at a longitude and
  latitude of $(0^{\circ},0^{\circ})$. Short radiative timescales result in steady-state $gh$
  profiles dominated by stellar forcing with a hot dayside and a cold
  nightside (see equation (\ref{2Dheq})), while the atmosphere relaxes to a
  constant $gh$ for long values of $\trad$. In contrast, the
  dependence of $gh$ on $\tdrag$ is weaker. 
The atmospheric circulation shifts from a zonal jet pattern
  at long $\trad$ and $\tdrag$ to day-to-night flow when either
  $\trad$ or $\tdrag$ is reduced, as explained in detail in
  \citet{Showman:2013p17094}.}
\label{fig-fullsolution}
\end{figure*}

\begin{figure*}
\includegraphics[width=7.0in]{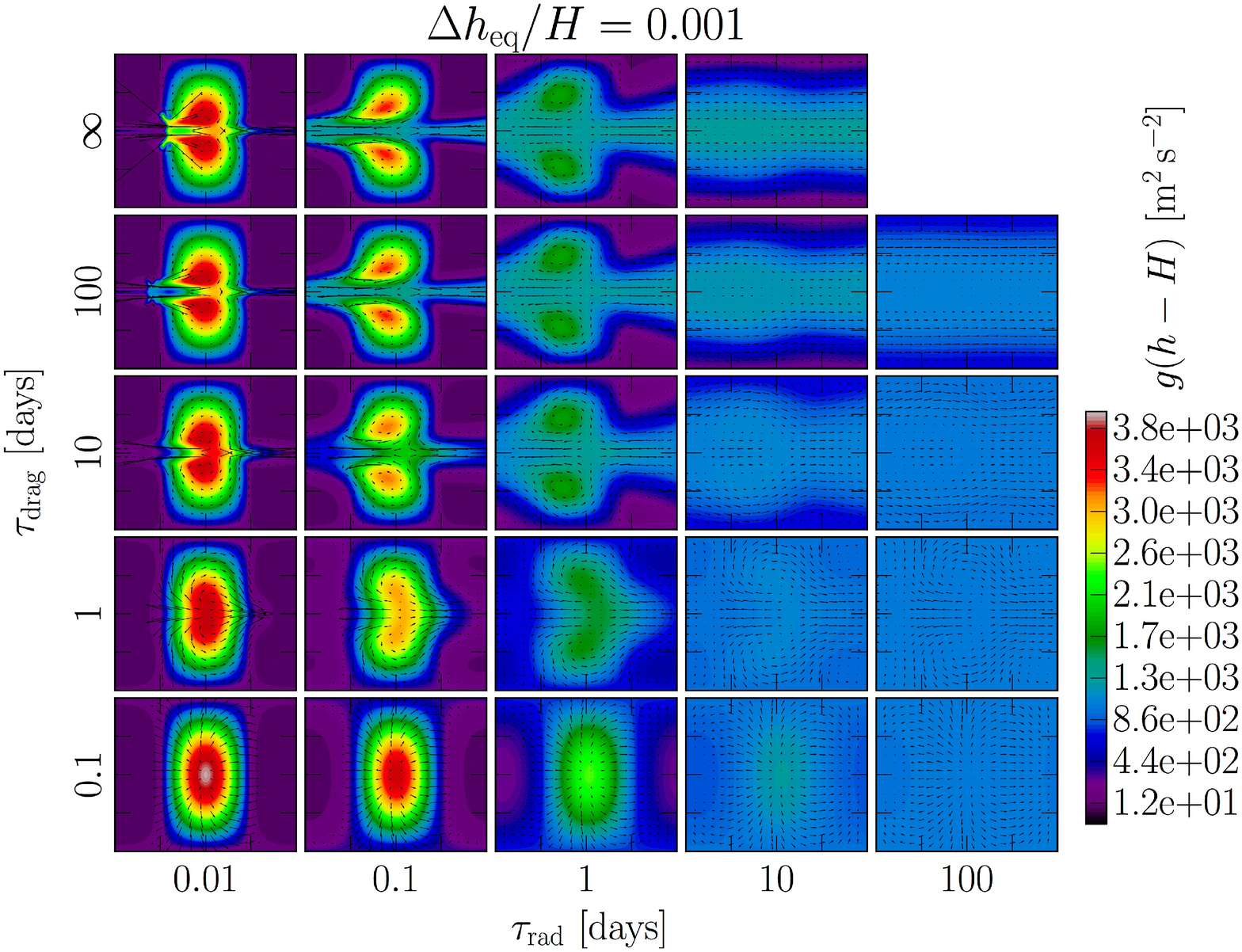}
\caption{%
Same as Figure \ref{fig-fullsolution}, but at a low forcing amplitude of $\forcing=0.001$. Note that the dynamic range of the $gh$-contour color-scale has been reduced by a factor of 1000 compared to Figure \ref{fig-fullsolution}. Wind speeds have also scaled down between figures. Comparing individual panels between Figures \ref{fig-fullsolution} and \ref{fig-fullsolution-0p001} reveals that even though the details of the solutions depend on $\forcing$, the characteristic value of the fractional day-night geopotential difference (normalized to the radiative equilibrium difference) remains largely unchanged.}
\label{fig-fullsolution-0p001}
\end{figure*}

Our equations apply to a two-layer system with a free upper
surface. For a derivation, see equations (3.37) and (3.38) in
\citet{Vallis:2006p12972}, which are identical to our equations with
the exception of the source terms, which we explain below. Note that
their momentum equation is written in terms of the height of the lower
layer (their $\eta_1$) above some reference level, whereas our
equations are expressed in terms of the upper layer thickness
($h$). Our momentum equation can be derived from theirs by noting that
isostasy in the lower layer requires $\nabla \eta_1 = -
\nabla h \rho\sub{upper}/\rho\sub{lower}$.

In local radiative equilibrium, the height field $h(\lambda, \phi,t) = h\sub{eq}(\lambda,\phi)$, with 

\begin{equation}
  h\sub{eq}= \left\{
  \begin{array}{l l}
    H+\Delta h\sub{eq}\cos{\lambda}\cos{\phi} & \quad \text{on the dayside}\\
    H & \quad \text{on the nightside,}\\
  \end{array} \right.
\label{2Dheq}
\end{equation}

\noindent where $H$ is the (flat) nightside thickness and $\Delta
h\sub{eq}$ is the difference in radiative-equilibrium thickness
between the substellar point and the nightside (see Figure
\ref{fig-sw}). The expression adopts a substellar point at
$(\lambda,\phi)=(0^{\circ},0^{\circ})$.  The planet is assumed to be
synchronously rotating so that $h\sub{eq}(\lambda,\phi)$ remains
stationary. Note that $h\sub{eq}$ represents the two-dimensional
height field set by local radiative equilibrium, whereas $\Delta
h\sub{eq}$ is its maximum deviation with respect to the layer
thickness $H$ at the nightside of the planet.

Equation~(\ref{2DShallowWaterB}) indicates that when the
upper layer is not in radiative equilibrium, mass will be
transferred between the layers, increasing or decreasing $h$. The
relaxation toward equilibrium occurs over a radiative timescale
$\tau\sub{rad}$---a free parameter in the model.  We can
understand this mass transfer over a radiative timescale in the
context of a 3D atmosphere. In a 3D context, $h$ represents the
amount of fluid having a specific entropy greater than a certain
reference value---i.e., $h$ is a proxy for the mass column above
an isentrope.  In regions that are heated to return to local
radiative equilibrium ($Q > 0$), fluid acquires entropy and rises
above the reference isentrope, increasing $h$. Likewise, in regions
that cool ($Q < 0$), fluid sinks below the reference isentrope, and $h$
decreases.

Mass transfer between the horizontally static lower layer and the
active upper layer will affect not only the local height of the upper
layer but also its momentum. The momentum advected with mass transfer
between layers is accounted for by $\textbf{R}$---the vertical transport term. In regions where
gas cools ($Q<0$), mass is locally transferred from the upper layer to
the lower layer. This process should not affect the specific momentum
of the upper layer, so $\textbf{R}=0$ when $Q<0$. In regions where gas
is heated ($Q>0$) mass is transferred from the lower layer into the
upper layer. The addition of mass with no horizontal velocity to the
upper layer should not affect its local column-integrated value of
$\textbf{v}h$. As explained by \citet{Showman:2011p12973}, one can
obtain an expression for $\partial(\textbf{v}h)/\partial t$ by adding
$\textbf{v}$ times the continuity equation to $h$ times the momentum
conservation equation. When $\partial(\textbf{v}h)/\partial t = 0$,
terms involving heating and cooling ($h\textbf{R}+\textbf{v}Q$) also
have to vanish, which yields $\textbf{R}=-\textbf{v}Q/h$ in regions of
heating. This expression for $\textbf{R}$ is also used in
\citet{Shell:2004p14317},
\citet{Showman:2010p12796, Showman:2011p12973}, and
\citet{Showman:2013p17094}.

Finally, we parameterize atmospheric drag
with Rayleigh friction, $-\textbf{v}/\tdrag$, where $\tdrag$ is a
specified characteristic drag timescale. \revd{Potential sources of atmospheric drag are hydrodynamic shocks and Lorentz-force drag. The latter is caused by ion-neutral collisions induced by magnetic deflection of thermally ionized alkali metals. In this work we keep a general prescription for drag that is proportional to the flow velocity, at the cost of missing details of the individual physical processes (e.g., Lorentz-force drag only affects the flow component moving orthogonally to the local planetary magnetic field and strongly depends on the local gas temperature \citep{Rauscher:2013p17720}).}

\begin{figure*}[ht]
\centering
\includegraphics[width=0.46\linewidth]{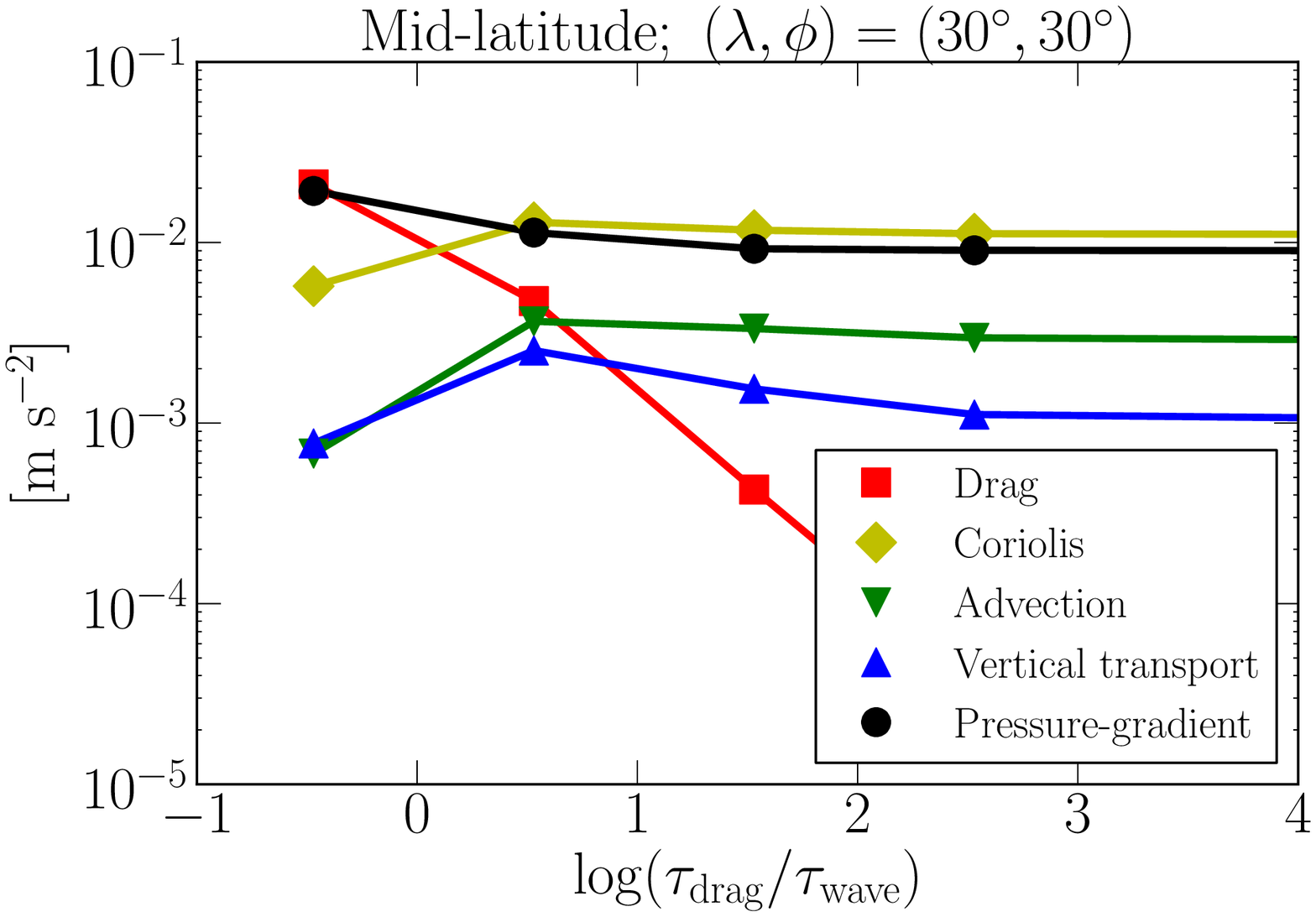}
\hspace{0.07\linewidth}
\includegraphics[width=0.46\linewidth]{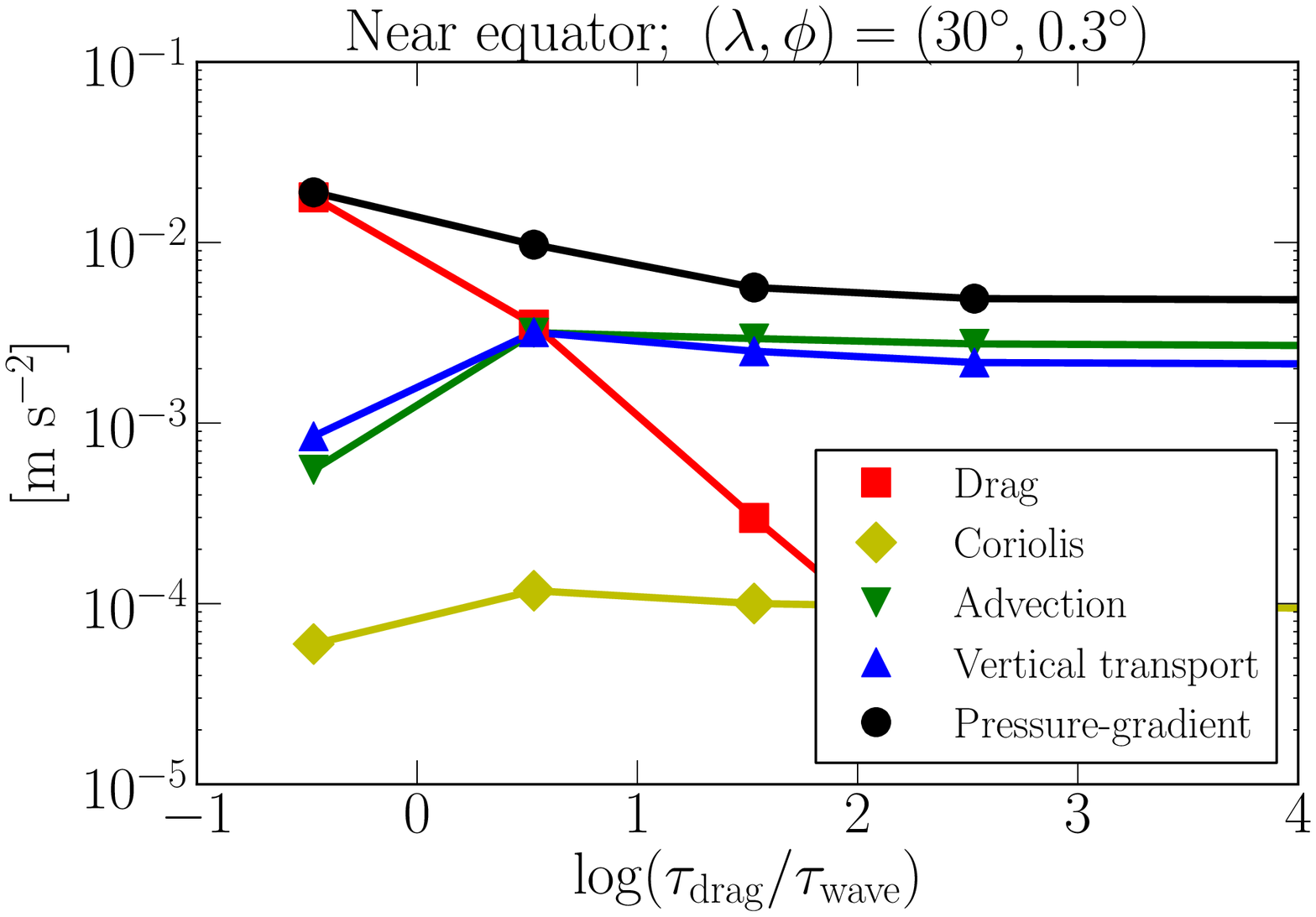}
\caption{%
Absolute values of the zonal components of individual terms in the momentum equation (\ref{2DShallowWaterA}) as a function of $\tdrag$. For both panels, $\forcing=1$ and $\trad$ is held constant at 0.1 days [$\log(\trad/\twave)=-0.5$]. The left panel is computed at a typical mid-latitude of the planet with coordinates $(\lambda,\phi)=(30^{\circ},30^{\circ})$, while the right panel is computed near the equator $(\lambda,\phi)=(30^{\circ},0^{\circ}.3)$. Far away from the equator (left panel), the primary balance in the momentum equation is between the pressure-gradient force and either the Coriolis force (weak-drag regime) or drag (drag-dominated regime). The transition from the weak-drag regime to the drag-dominated regime in the momentum balance occurs when $\tdrag \sim 1/f$ (equation (\ref{eq_white1})). Near the equator (right panel) the primary term balancing the pressure-gradient force is either advection (together with the vertical transport term $\textbf{R}$) or drag.}
\label{fig-mb}
\end{figure*}

\begin{figure*}[ht]
\centering
\includegraphics[width=0.46\linewidth]{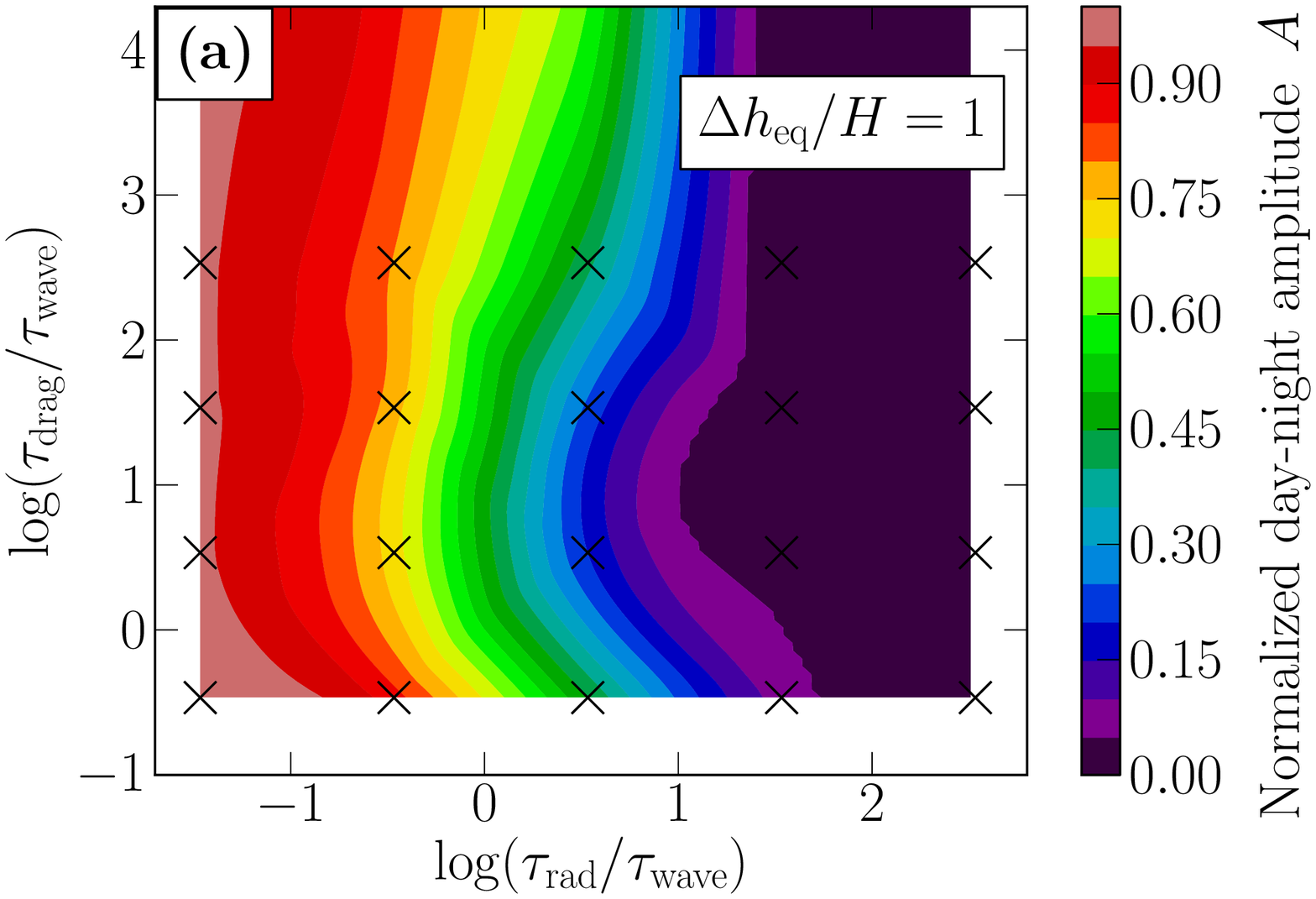}
\hspace{0.07\linewidth}
\includegraphics[width=0.46\linewidth]{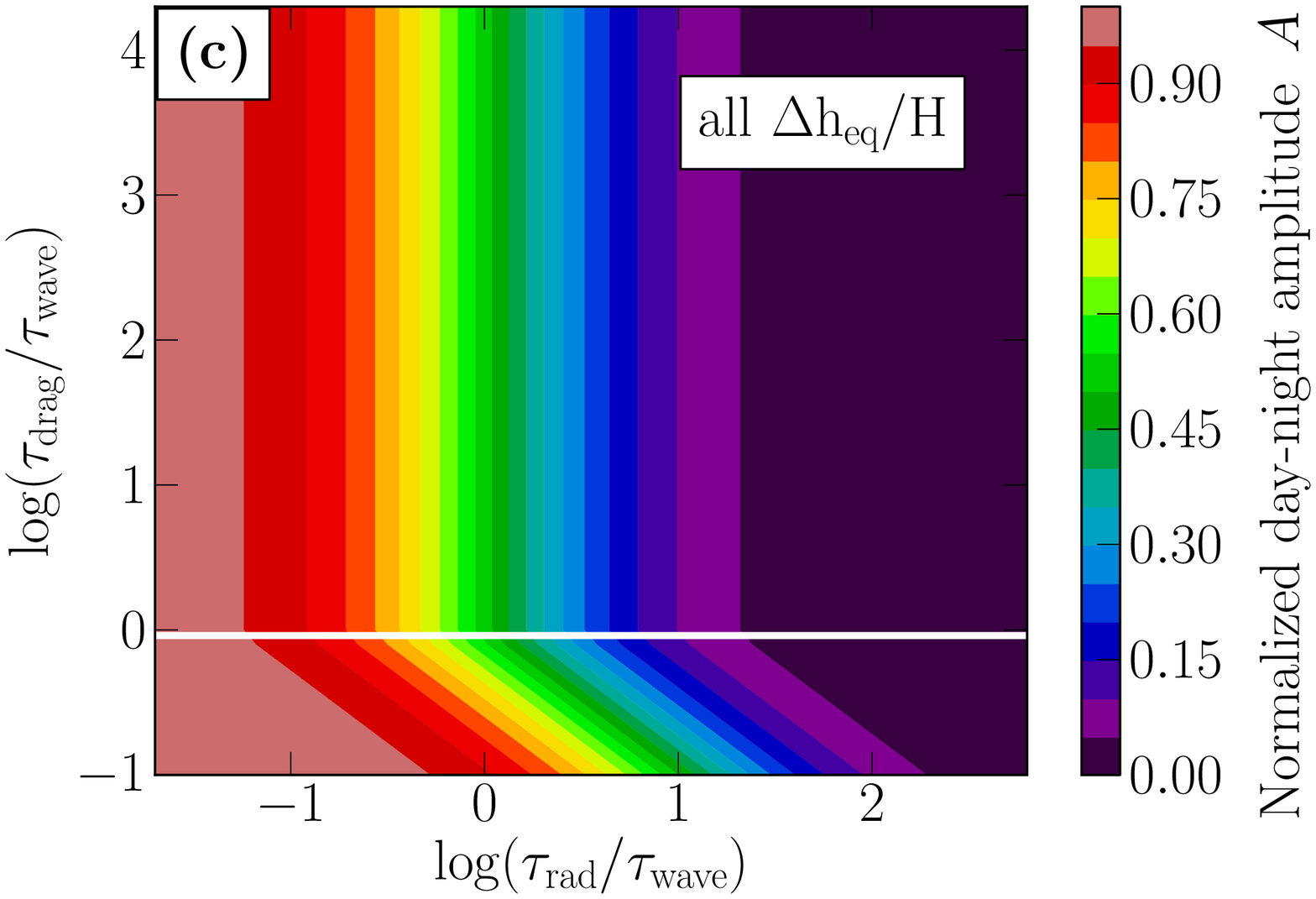}
\includegraphics[width=0.46\linewidth]{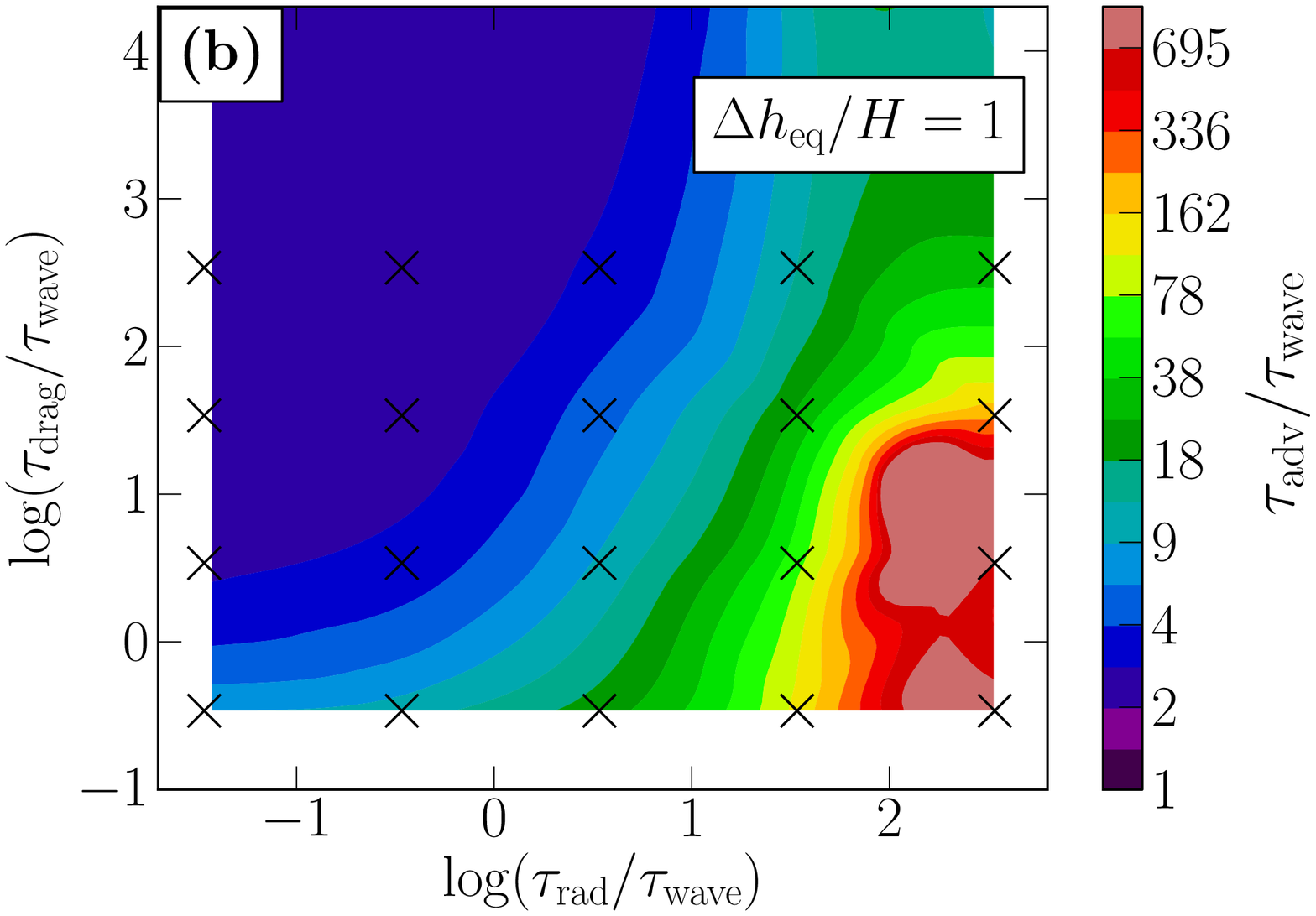}
\hspace{0.07\linewidth}
\includegraphics[width=0.46\linewidth]{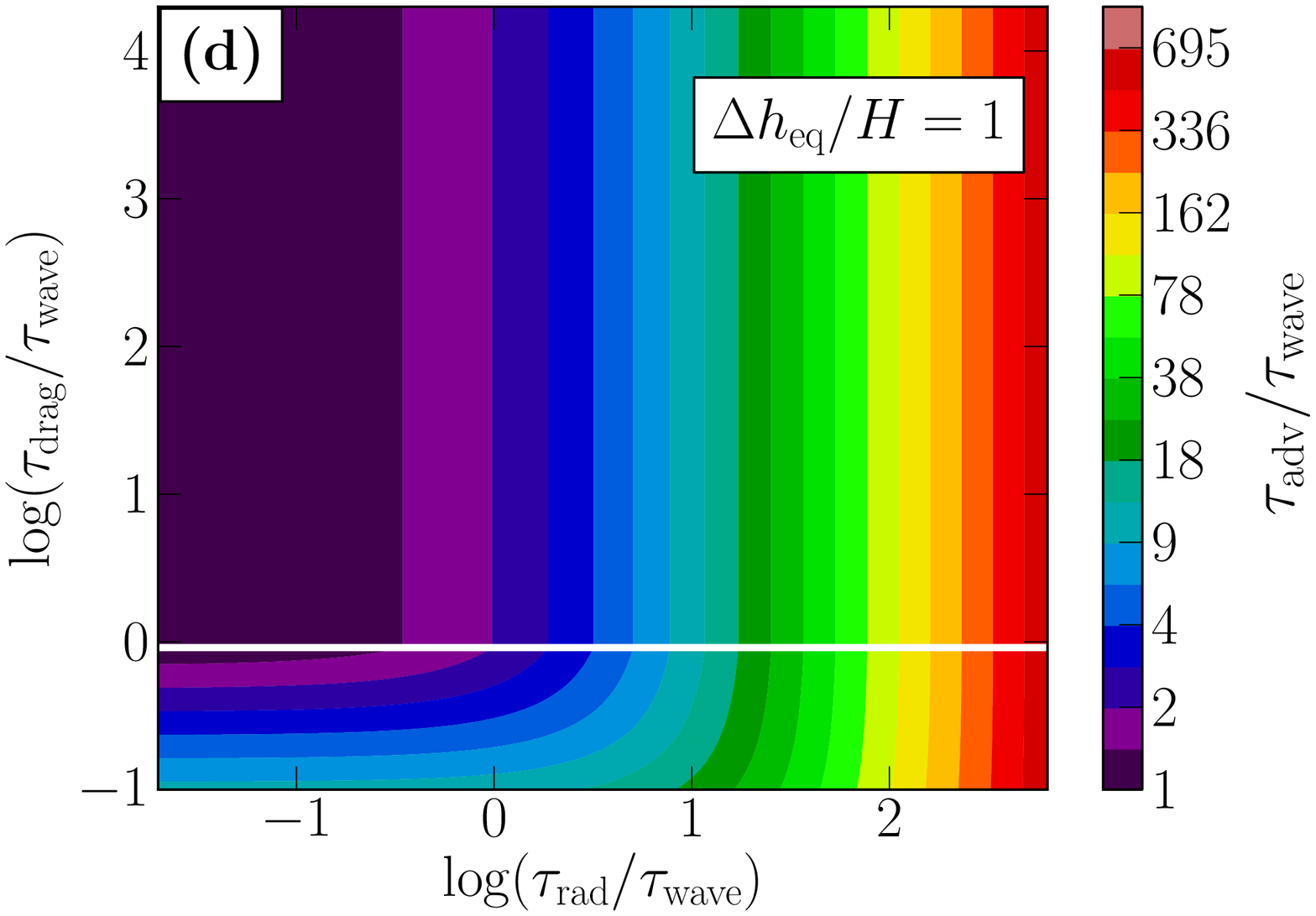}

\caption{\textbf{Panel (a):} contours of day-night
  height amplitude $A$ (defined by equations (\ref{A_definition})--(\ref{A_integrated})) as a function of $\trad$ and $\tdrag$ from our full numerical simulations with $\forcing=1$. Values range from $A=0$, which corresponds to a planet with a constant height field, to $A=1$, which corresponds to a planet with $h=h\sub{eq}$ everywhere, i.e., zero heat redistribution. Simulations were carried out for a matrix of $\trad$ and $\tdrag$ values as described in Figure \ref{fig-fullsolution}. These runs are marked by the $\times$ symbol; intervening values were calculated with a cubic-spline interpolation. In general terms, when $\trad$ is short compared to $\twave$, atmospheres have a small heat redistribution efficiency.  %
  \textbf{Panel (b):} contours of $\tadv/\twave \equiv \sqrt{gH}/U$ from these same simulations as a
  function of $\trad$ and $\tdrag$ with $\forcing=1$. Here $U$ is the RMS value of
  atmospheric winds (both zonal and meridional) averaged over the
  entire planet, and $\sqrt{gH}$ is the gravity wave speed. Throughout
  the sampled space $\tadv/\twave > 1$, with a minimum value of $2.1$,
  colored purple. In the global mean, gravity waves always travel faster than winds. 
\textbf{Panel (c):} contours of day-night geopotential
  amplitude $A$ as a function of $\trad$ and $\tdrag$ as predicted by
  our analytical theory (equation (\ref{eq-A-combined})). Depending on the strength of atmospheric drag, different
  terms are being balanced in the momentum
  equation, yielding two distinct regions in the plot, separated by
  the solid white line, defined by equation (\ref{eq_white1}). Drag is irrelevant in the upper region, but
  plays a significant role in the lower one. The simple timescale
  comparisons of our analytical theory broadly reproduce the results
  of the full numerical shallow-water model shown in panel (a). The scaling theory 
predicts that $A$-contours are independent of $\forcing$.
 \textbf{Panel (d):} contours of $\tadv/\twave$ as a function of $\trad$ and $\tdrag$ as predicted by our analytical theory (equation (\ref{eq_oom3b}) for the drag-dominated regime below the white line and equation (\ref{eq_oom5b}) for the weak drag regime above the white line). In contrast to $A$, contour values of $\tadv/\twave$ depend on $\forcing$.}
\label{fig-A-model}
\end{figure*}

We solve equations (\ref{2DShallowWaterA}) and (\ref{2DShallowWaterB})
in global, spherical geometry with the following parameter choices. We
choose $g$ and $H$ such that the gravity wave speed is $\sqrt{gH} =
(10 \, {\rm m} \, {\rm s}^{-2} \times 400 \, {\rm km})^{1/2} = 2$ km
s$^{-1}$.\footnote{Our model is specified by the product $gH$ and does
  not require separate specification of $g$ and $H$, as can be seen by
  multiplying equation~(\ref{2DShallowWaterB}) by $g$. \revd{Gravity waves in our shallow water model are a proxy for internal gravity waves that exist in stratified atmospheres. In contrast to shallow water---where waves propagate exclusively in the plane perpendicular to gravity---gravity waves in a stratified atmosphere can propagate in any direction. Nevertheless, the latitudinal forcing on hot Jupiters will primarily excite horizontally traveling gravity waves whose propagation speeds will be similar to $\sqrt{g_{\rm actual} H_{\rm atm}}$, where $g_{\rm actual}
  \approx 10$ m s$^{-2}$ is the full gravitational acceleration (not
  the reduced gravity) and $H_{\rm atm} \approx 400$ km is the
  atmospheric pressure scale height. Our nominal
  value for $gH$ is chosen to match the propagation speed of the gravest mode of zonal gravity waves at the photosphere of hot Jupiters.} Although it may be tempting
  to do so, we do not think of $g$ as approximating $g_{\rm actual}$
  (since the reduced gravity is conceptually distinct from and
  generally differs from the full gravity), nor do we think of $H$ as
  approximating $H_{\rm atm}$, since $H$ is instead supposed to be a
  proxy for the thickness of material above an isentrope. \revd{Vertically
propagating gravity waves are not captured by our model, but seem to be of lesser importance for the dynamics as they are not being excited by the longitudinal heating gradient---the driver of the system.}}
For our fiducial hot Jupiter model,
we set the relative forcing amplitude $\Delta h\sub{eq}/H=1$, implying
that in radiative equilibrium temperature differences between day- and
nightside vary by order unity. We will vary $\Delta h\sub{eq}/H$ down
to $0.001$ to understand dynamical mechanisms and to verify that our
theoretical predictions are valid in the linear forcing limit.  Our
values for the rotation frequency $\Omega=3.2\e{-5}$ s$^{-1}$ and a
planetary radius of $a = 8.2\e{7}$ m are similar to those of HD
189733b. The characteristic wave travel timescale is
\begin{equation}
\label{eq-twave}
\twave \sim {L\over \sqrt{gH}} \sim
0.3\,\, \rm{Earth\,\,days}, 
\end{equation}
\noindent where $L$ is the characteristic horizontal length scale
of the flow.  For typical hot Jupiters, this is comparable to
$L_{\rm eq}\equiv (\sqrt{gH} a/2\Omega)^{1/2} \approx 5.1 \e{7}$m,
the equatorial Rossby deformation radius---a natural length scale
that results from the interaction of buoyancy forces and Coriolis forces
in planetary atmospheres.
Note that for a typical hot Jupiter, $L$ and $L_{\rm eq}$
happen to be of the same order as the planetary radius $a$ \citep{Showman:2002p12764, Menou:2003p17826,
Showman:2011p12973}.  We explore
how the solution depends on the characteristic damping timescales
$\trad$ and $\tdrag$, which are free parameters in the model. In
contrast, the characteristic time over which a gas parcel is advected
over a global scale, $\tadv$, depends on the resulting wind profile
and cannot be independently varied.

We solve equations (\ref{2DShallowWaterA}) and (\ref{2DShallowWaterB})
in spherical geometry with the Spectral Transform Shallow Water Model
(STSWM) of \citet{Hack:1992p0001}. The equations are integrated using
a spectral truncation of T170, corresponding to a resolution of
$0^{\circ}.7$ in longitude and latitude (i.e., a 512 $\times$ 256 grid
in longitude and latitude). A $\nabla^{6}$ hyperviscosity is applied
to each of the dynamical variables to maintain numerical
stability. The code adopts the leapfrog time-stepping scheme and
applies an Asselin filter at each time step to suppress the
computational mode. Models are integrated from an initially flat layer
at rest until a steady state is reached. This system has been shown
to be insensitive to initial conditions by \citet{Liu:2013p17827}. The models
described here include those presented in \citet{Showman:2013p17094},
as well as additional models that we have performed for the present
analysis.

In summary, the model contains three main input parameters: $\trad$,
$\tdrag$, and $\forcing$. Our main goal in Section \ref{sec-numerical-solutions} is
to determine the dependence of the equilibrated fractional day-night
height difference on these parameters. In Section \ref{sec-theory} we
develop a simple analytic scaling theory that reproduces trends found
in the numerical model.

\section{NUMERICAL SOLUTIONS}
\label{sec-numerical-solutions}

\subsection{Basic Behavior of the Solutions}

As discussed in Section \ref{sec-intro}, observations indicate that as
the stellar insolation increases, atmospheres transition from having
small longitudinal temperature variations to having large day-night
temperature contrasts. Our model solutions capture this transition, as
shown in Figure \ref{fig-fullsolution}.  There, we plot the difference between
the steady-state geopotential ($gh$) and the nightside geopotential 
at radiative equilibrium ($gH$) for twenty-five models performed at high
amplitude ($\Delta h_{\rm eq}/H=1$) over a complete grid in $\trad$
and $\tdrag$.  Models are shown for all possible combinations of
0.01, 0.1, 1, 10, and 100 Earth days in $\trad$ and 0.1, 1, 10, 100,
and $\infty$ days in $\tdrag$.  In the context of a 3D atmosphere, $h$
represents the mass column above a reference isentrope; large $h$
represents more material at high specific entropy (high temperature on
an isobar). The stellar insolation is varied in the model by adjusting
the damping timescales $\trad$ and $\tdrag$. Generally, the higher the
stellar flux (as measured by $T\sub{eq}$), the lower $\trad$ will be. 
We will quantify the dependence of $\trad$ on $T\sub{eq}$ in Section \ref{sec-observations}.

The models in Figure~\ref{fig-fullsolution} capture major transitions
in both the structure of the flow and the amplitude of the day-night
thickness contrast.  When $\trad$ is longer than one Earth-day,
longitudinal gradients of $gh$ are small.  If $\tdrag$ is also long
compared to a day, the circulation primarily consists of
east-west-aligned (zonal) flows varying little in longitude (upper
right corner of Figure~\ref{fig-fullsolution}).  Despite the lack of
longitudinal variation, such models exhibit an equator-pole gradient
in $gh$, albeit with an amplitude that remains small compared to the
radiative-equilibrium gradient.  When $\trad$ is long but $\tdrag$ is
short, winds flow from the dayside to the nightside over both the
eastern and western hemispheres, and $gh$ varies little in either
longitude or latitude (lower right corner of
Figure~\ref{fig-fullsolution}). Intermediate values of $\trad$ (e.g.,
$\sim$1 day; middle column of Figure~\ref{fig-fullsolution}) lead to
flows with greater day-night temperature differences and significant
dynamical structure, including zonal-mean zonal winds that are eastward at the equator
(i.e., equatorial superrotation). In contrast, when $\trad$ is short
(left column of Figure~\ref{fig-fullsolution}),
the geopotential amplitude and morphology closely match the
radiative forcing profile: a spherical bulge on the hot dayside
and a flat, cold nightside (see equation \ref{2Dheq}).  The circulation
consists of strong airflow from day to night along both
terminators. \citet{Showman:2011p12973} and \citet{Showman:2013p17094} showed that
much of the wind behavior in Figure~\ref{fig-fullsolution} (and in
many published 3D global circulation models of hot Jupiters)
can be understood in terms of the interaction of standing,
planetary-scale waves with the mean flow.

Many of the characteristics of the full solution can be understood by
studying the model under weak forcing ($\forcing \ll 1$).  In this
limit, the day-night variations in $h$ are much smaller than $H$, and
terms in the shallow water equations exhibit their linear
response. For example, the term $\nabla\cdot({\bf v}h)$ in the
continuity equation will behave approximately as $H\nabla\cdot{\bf
  v}$. The balance between $H\nabla\cdot{\bf v}$ and $Q$ in the
continuity equation is linear. If the balance in the momentum equation
is also linear, then wind speeds ${\bf v}$ and amplitudes of $h$-variation should scale linearly with the forcing amplitude
$\forcing$. In Figure \ref{fig-fullsolution-0p001} we present
solutions of the model forced at the low amplitude of $\forcing=0.001$
for the same values for $\trad$ and $\tdrag$ as in Figure
\ref{fig-fullsolution}. Note that contour values for $g(h-H)$ have
been scaled down by a factor of 1000 from those in Figure
\ref{fig-fullsolution}. At this low amplitude, the system responds
linearly for most of parameter space.   
A comparison of Figures~\ref{fig-fullsolution-0p001} and 
\ref{fig-fullsolution} demonstrates that the low-amplitude
  behavior is extremely similar to the high-amplitude behavior when
  $\tau_{\rm drag}$ is short, but differs when $\tau_{\rm drag}$ is
  long.  The amplitude dependence under weak-drag conditions is greatest
  when $\tau_{\rm rad}$ is short: at low amplitude, the mid-and-high
  latitudes are close to radiative equilibrium, whereas the equator
  exhibits almost no longitudinal variations in $gh$.

\subsection{Physical Explanation for Forcing Amplitude Dependence}

We now
proceed to examine the reasons for these amplitude differences.
Consider the principal
force balances that determine the model solutions in the linear limit
as a function of both drag and latitude.

For sufficiently strong drag ($\tdrag \lesssim 1$ day), the balance in the momentum equation is primarily between the pressure-gradient force---which drives the flow---and drag. This is a linear balance, and because the term balance in the continuity equation is likewise linear, we expect the $h$-field to scale with $\forcing$. This is indeed the case, as can be appreciated by the similarity of the lower panels of Figures \ref{fig-fullsolution} and \ref{fig-fullsolution-0p001}.%

When drag is reduced to the point where it becomes negligible ($\tdrag
\gtrsim 1$ day), other terms in the momentum equation have to balance
the pressure-gradient force. Which term plays the dominant role
depends on the latitude of the planet. This is evident in Figure
\ref{fig-fullsolution-0p001}, where solutions in the upper left corner
have a flat $h$-field at the equatorial region, while the $h$-field
has a day-night amplitude that approaches radiative equilibrium at
high latitudes. For $\forcing=0.001$, winds are weak, and away from
the equator, the Rossby number $Ro=U/fL \sim 0.001 \ll 1$, where $U$
is a characteristic horizontal wind speed. As a result, the primary
force balance away from the equator is between the Coriolis force and
the pressure-gradient force. This force balance is linear. Because the
continuity equation is also linear, both $h$ and ${\bf v}$ should
scale with $\forcing$ at mid-latitudes. Comparing the upper rows of
Figures \ref{fig-fullsolution} and \ref{fig-fullsolution-0p001}
confirms that $h$-fields away from the equator scale with forcing
amplitude in the weak-drag limit. Nevertheless, as the forcing
amplitude is raised, wind speeds increase until $Ro \sim 1$ when
$\forcing = 1$. Therefore, the advective term becomes comparable to
the Coriolis term at high amplitudes. This results in differences, but
no fundamental changes, in the flow structure at high latitudes. The
linear dynamics in the weak drag regime are described in more detail in
Appendix C of \citet{Showman:2011p12973}.

In the weak-drag limit, why does the equator exhibit large fractional height
variations at large forcing amplitude but only small fractional height variations at small forcing amplitude?
At the equator, the Coriolis force vanishes ($Ro \gg 1$) and, if drag is weak, the force balance is between the pressure-gradient force and advection.  
This is an inherently non-linear balance because advection scales with the square of the velocity. 
Thus there is no linear limit for the dynamical behavior at the equator, and $h$ will not scale
linearly with $\forcing$. At $\forcing=1$, the advection term is comparable to the Coriolis term at mid-latitudes. As a result, pressure (height) gradients remain as large at the equator as they are at mid-latitudes as can be appreciated in the upper rows of Figure \ref{fig-fullsolution}. However, as the forcing amplitude is reduced to $\forcing=0.001$, the advection term diminishes at a quadratic rate. To maintain balance with advection, the pressure-gradient force must also weaken as $(\forcing)^2$. Thus the height field becomes flat near the equator, as evidenced in the upper rows of Figure \ref{fig-fullsolution-0p001}.

In Figure \ref{fig-mb} we plot the magnitudes of the zonal components of all terms in the momentum equation as a function of $\tdrag$. The term balances discussed above are apparent. Figure \ref{fig-mb} is computed for $\forcing=1$, with $\trad$ held constant at 0.1 days, while $\tdrag$ is varied in the abscissa. Note that Figure \ref{fig-mb} normalizes $\tdrag$ with $\twave$---the (constant) wave travel time (equation (\ref{eq-twave})). The relevance of $\twave$ will
be explained in Section \ref{sec-theory}; from this point onward, we
will express timescales in terms of $\twave$.  The left panel plots the terms at a typical mid-latitude with coordinates ($\lambda$, $\phi$)=($30^{\circ}$, $30^{\circ}$), while the right panel is for a point near the equator ($\lambda$, $\phi$)=($30^{\circ}$, $0^{\circ}.3$). As noted before, the pressure-gradient force away from the equator is balanced primarily against either the Coriolis force or drag. Near the equator, advection and vertical transport balance the pressure gradient when drag is weak.

\subsection{Metric for the Day-Night Contrast}

To compare our model solutions to the observed fractional flux
variations of extrasolar planets, as well as to theoretical
predictions, we need a measure similar to
$A\sub{obs}$ (see Figure \ref{fig-obs-nomodel}). Our proxy for 
flux variations will be a day-night 
height difference $A$ representative for the entire planet.
We thus reduce each panel in Figures \ref{fig-fullsolution} and \ref{fig-fullsolution-0p001} 
to a single $A$, which we compute as follows. We start by evaluating the root-mean-square
variations of $h$ over circles of constant latitude, and normalize
them to the values at radiative equilibrium:

\begin{equation}
	A(\phi)=\left\{\frac{\int_{-\pi}^{\pi}\left[h(\lambda,\phi)-\bar{h}(\phi)\right]^2 d\lambda}{\int_{-\pi}^{\pi}\left[h\sub{eq}(\lambda,\phi)-\bar{h}(\phi)\right]^2 d\lambda}\right\}^{1/2}, 
\label{A_definition}
\end{equation}

\noindent where $\bar{h}(\phi)$ is the zonally averaged height at a given latitude
\begin{equation}
\bar{h}(\phi)=\frac{1}{2\pi}\int_{-\pi}^{\pi} h(\lambda,\phi) \,d\lambda\,.
\label{hbar_definition}
\end{equation}

\noindent We then average $A(\phi)$ over a $60^{\circ}$ band centered at the equator
\begin{equation}
	A=\frac{3}{\pi}\int_{-\pi/6}^{\pi/6}A(\phi)  \,d\phi\,.
\label{A_integrated}
\end{equation}

\noindent We find that for bands of width $\gtrsim 60^{\circ}$, $A$ becomes insensitive to the range in latitudes used for averaging. As defined, $A$ can vary from $0$, when $h(\lambda,\phi,t)=\bar{h}(\phi)$ everywhere (corresponding to an atmosphere without longitudinal height variations), to $1$, when the height equals that imposed by radiative forcing, $h(\lambda,\phi,t)=h\sub{eq}(\lambda,\phi)$. %

Figure \ref{fig-A-model}(a) shows how model values of $A$ depend on the
choice of damping timescales $\trad$ and $\tdrag$ (which have been
normalized to $\twave$) for high-amplitude models with $\Delta h_{\rm
  eq}/H = 1$.
Models with short radiative time constants exhibit large 
fractional day-night differences.  When the drag timescale is
short ($\tdrag/\twave < 1$), friction in the atmosphere starts to play
a more important role in controlling the day-night height
difference. Figure \ref{fig-A-model}(b) shows contours of $\tadv/\twave
\equiv \sqrt{gH}/U$ as a function of $\trad$ and $\tdrag$. Here $U$ is
the RMS-value of atmospheric winds (both zonal and meridional)
averaged over the entire planet, and $\sqrt{gH}$ is the gravity wave
speed. Throughout the sampled space $\tadv/\twave > 1$, with a minimum
value of $2.1$, colored purple. Since $\twave < \tadv$ everywhere, 
the characteristic global-mean speed of gravity waves is always 
faster than the characteristic global-mean wind speed (however, note
that, at the highest forcing amplitude, this is not always true
locally everywhere over the globe).\footnote{\revd{Supercritical flows with $U > \sqrt{gH}$ occur near the day-night terminator in the long $\tdrag$ and short $\trad$ limit (upper-left panels in Figure \ref{fig-fullsolution}). When supercritical flow rams into slower moving fluid, hydraulic jumps develop which convert some kinetic energy into heat \citep{Johnson:1997p17920}. We do not account for this source of heating as it is likely to be modest and is only present for solutions where $A$ is already $\sim$$1$. Hydraulic jumps can also occur for supercritical flows in a stratified atmosphere. These are distinct from acoustic shocks that can develop in supersonic flow.
The importance of acoustic shocks in modifying the photospheric temperature profile of hot Jupiters is still an open question, as most global simulations (including ours) do not capture the relevant physics, i.e., sound waves, and/or lack sufficient spatial resolution \citep{Li:2010p14315}.}}

We can explain the qualitative behavior of these numerical solutions
with an analytic theory in which we substitute dominant terms in the mass and
momentum conservation equations with their order-of-magnitude counterparts. These analytical predictions, derived
in Section \ref{sec-theory}, are showcased in Figures
\ref{fig-A-model}(c) and (d)---which are a reasonable
match to Figures \ref{fig-A-model}(a) and (b) showing the
results of the numerical model. Two behaviors become apparent: above
the white line of Figure \ref{fig-A-model}(c),
the contours are vertical, indicating that atmospheric drag
does not affect the day-night temperature variation, a prediction that is in agreement with Figure \ref{fig-mb}.  Below the white line, both the radiative and drag timescales affect $A$. Our scaling theory will confirm that $\twave$ plays a central role in controlling the heat redistribution efficiency.

\begin{figure}[ht]
\centering
\includegraphics[width=\linewidth]{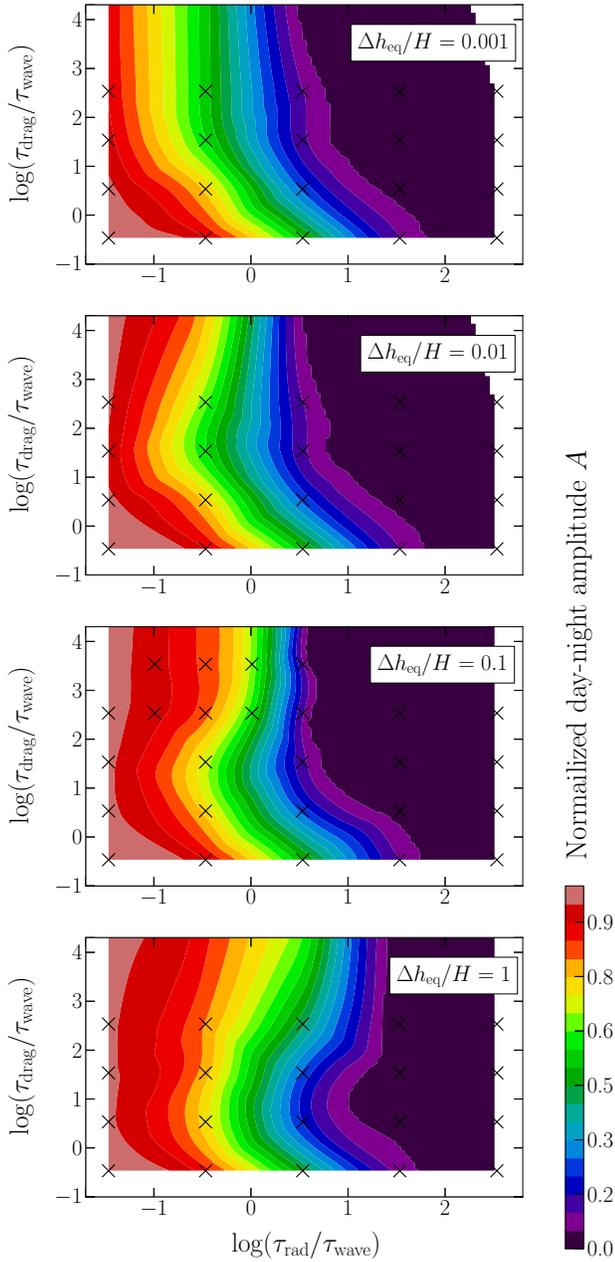}
\caption{%
Contours of normalized day-night amplitude $A$ (equations (\ref{A_definition})--(\ref{A_integrated})) from our full shallow water simulations as a function of $\trad$ and $\tdrag$. Each panel was computed for a different $\forcing$, ranging from $0.001$ (top panel) to $1$ (bottom panel). The lowermost panel is identical to Figure \ref{fig-A-model}(a) and is repeated to facilitate a direct comparison with the other panels. Over the three orders of magnitude in $\forcing$ shown, the morphology of $A$ seems to remain roughly unchanged, with nearly vertical $A$-contours in the upper half of each panel and slanted $A$-contours for the lower half, where the drag force dominates the momentum equation. All panels compare well to our analytical scaling theory, shown in Figure \ref{fig-A-model}(c).}
\label{fig-global-forcing}
\end{figure}

\begin{figure*}
\centering
\includegraphics[width=7.0in]{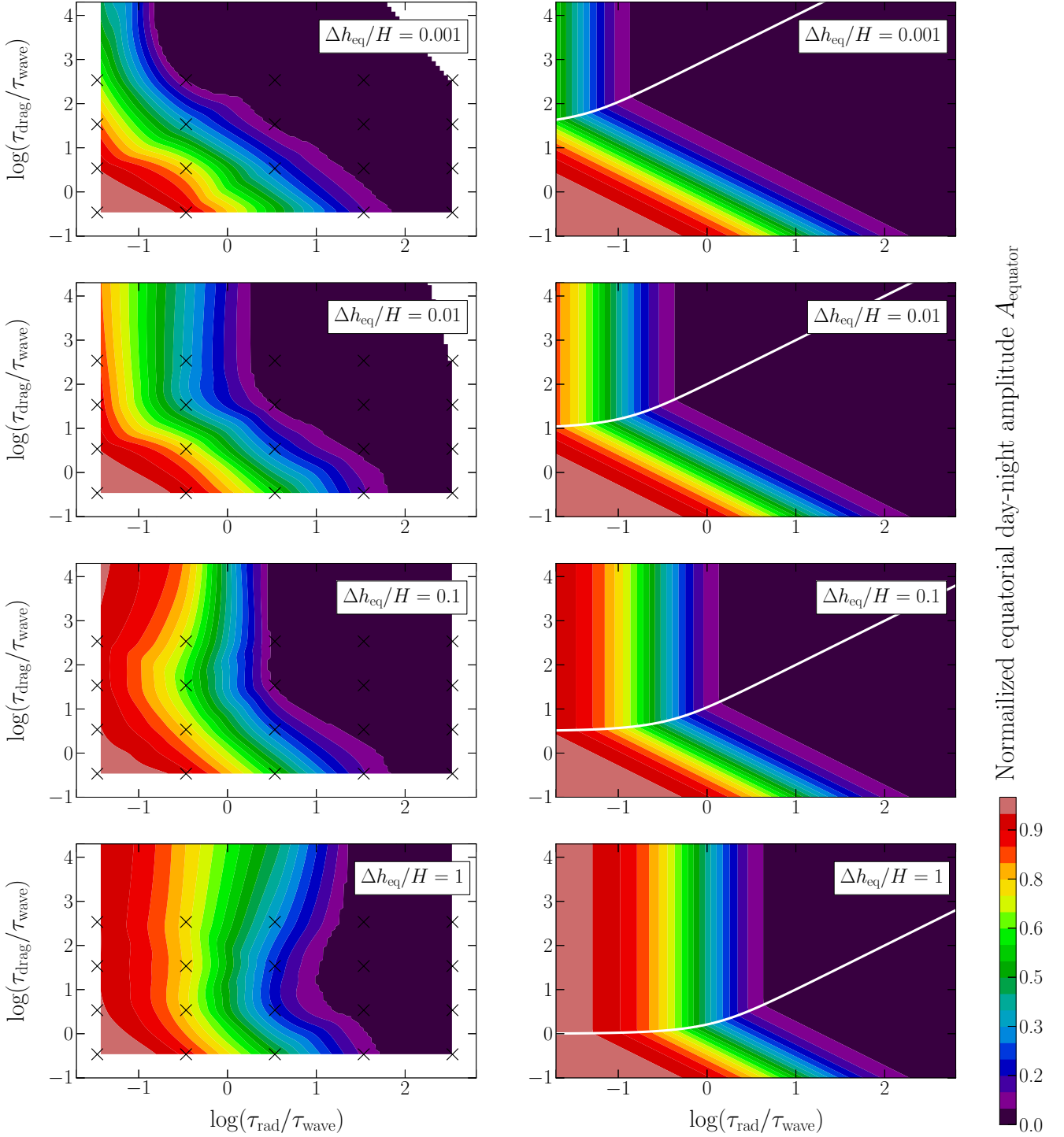}
\caption{%
\textbf{Left-hand panels:} same as Figure \ref{fig-global-forcing}, but for $A_{\rm{equator}}$, the normalized day-night amplitude from our full shallow-water simulations evaluated only at the planetary equator. For $\forcing=1$, $A_{\rm{equator}}$ contours are similar to those of $A$ (shown in the lowermost panel of Figure \ref{fig-global-forcing}). But as $\forcing$ is reduced, more of the $\trad$, $\tdrag$ parameter space becomes drag-dominated, characterized by slanted $A_{\rm{equator}}$ contours. At $\forcing=0.001$ (uppermost panel), most of the parameter space is drag dominated and the height field becomes flat at the equator when $\tdrag \to \infty$ (as can be appreciated directly from model solutions shown in the upper rows of Figure \ref{fig-fullsolution-0p001}). \textbf{Right-hand panels:} normalized day-night equatorial amplitude $A_{\rm{equator}}$ as predicted from our scaling theory. The white line marks the transition from the low-drag regime (where $A_{\rm{equator}}$ is given by equation (\ref{eq_oom4})) to the drag-dominated regime (where $A_{\rm{equator}}$ is given by equation (\ref{eq_oom3})). Our theory compares well with the numerical shallow water solutions shown in the left panels of this figure.}
\label{fig-equator-forcing}
\end{figure*}

In Figure \ref{fig-global-forcing} we show the variation of $A$ (as defined in
equations \ref{A_definition}--\ref{A_integrated}) with forcing
amplitude $\forcing$. The morphology of $A$ is largely independent of forcing strength.
The two upper panels show the solution close to the linear
limit, where we expect $A$ to be independent of $\forcing$.\footnote{For low $\forcing$, the non-linear equatorial region---defined by $Ro \gtrsim 1$---is thin and does not significantly contribute to the integral that makes up $A$ in equation (\ref{A_integrated}).}
As $\forcing \rightarrow 1$, $Ro$ increases to order-unity values at mid-latitudes. Nevertheless, the Coriolis force remains comparable to or greater than advection in regions away from the equator (see the left panel of Figure \ref{fig-mb}).
Thus, even for a large forcing amplitude, $A$ remains largely independent of $\forcing$, because the primary force balance is close to linear. 
In contrast, the day-night difference evaluated solely at the equator, $A_{\rm{equator}} \equiv
A(\phi=0)$, will depend on $\forcing$. In the left
panels of Figure~\ref{fig-equator-forcing} we show $A_{\rm{equator}}$-contours from our full numerical simulations as a function of $\trad$, $\tdrag$, and $\forcing$. The right panels of Figure \ref{fig-equator-forcing} show the
$A_{\rm{equator}}$-contours predicted by our scaling theory (Section \ref{sec-theory}). 
In the strong-drag regime
(region below the white line), the behavior of $A_{\rm{equator}}$ in
Figure~\ref{fig-equator-forcing} is relatively independent of
amplitude, because the balance between drag and pressure-gradient forces is linear. 
In the weak-drag regime, the force balance is between advection and the pressure-gradient force. For this non-linear balance there is no linear limit for the dynamical behavior at the equator, and $A_{\rm{equator}}$ depends on the forcing amplitude.

\begin{figure}[h!]
\includegraphics[width=\linewidth]{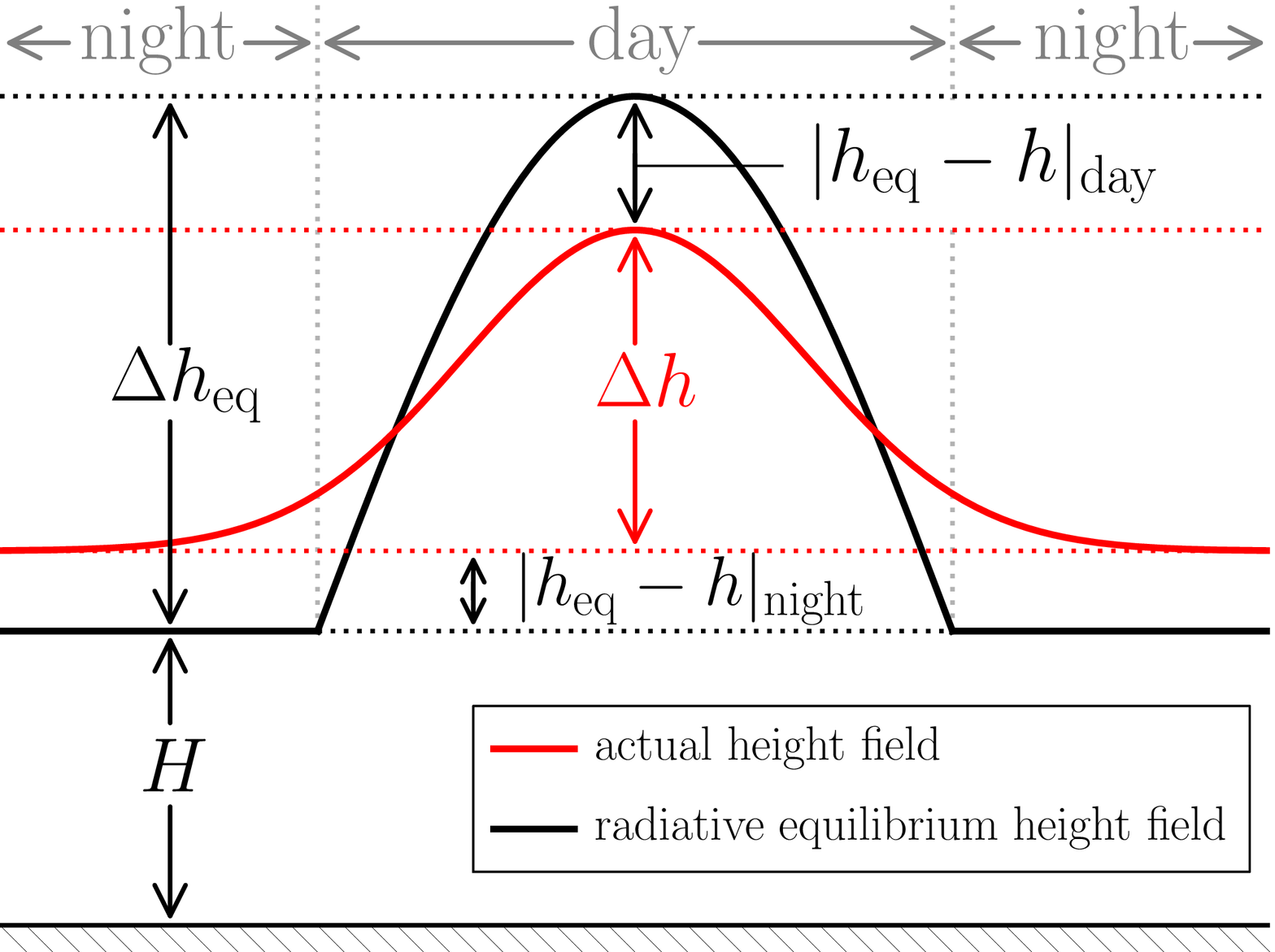}
\caption{%
Simplified diagram of the upper layer of the shallow water model. We show this diagram as an aid to understanding equation (\ref{eq-daynight}). The interface between the upper and lower model layers is drawn as a flat floor (cf. Figure \ref{fig-sw}). Both the actual height field $h$ (red solid line) and the radiative equilibrium height field $h\sub{eq}$ (black solid line) are now measured with respect to this floor. We define a characteristic difference between $h\sub{eq}$ and $h$ on the dayside ($|h_{\rm eq} - h|_{\rm{day}}$) and on the nightside ($|h_{\rm eq} - h|_{\rm{night}}$) of the planet.}
\label{fig-sw2}
\end{figure}

\section{ANALYTIC THEORY FOR DAY-NIGHT DIFFERENCES}
\label{sec-theory}

Here we obtain an approximate analytic theory for the day-night
thickness differences and wind speeds in the equilibrated steady
states. Our full numerical nonlinear solutions exhibit
steady behavior, so the partial time derivatives in both
the continuity and momentum equations can be neglected. 
The mass conservation equation~(\ref{2DShallowWaterB}) can thus be approximated as
\begin{equation}
	h (\nabla \cdot \textbf{v}) + \textbf{v} \cdot \nabla h \sim \frac{h\sub{eq}-h}{\trad}. 
\label{mass-cont}
\end{equation} 
\noindent 
On the right side, the quantity $h_{\rm eq} - h$ gives the difference between the local radiative-equilibrium
 height field and the local height field (this difference is generally positive on the dayside
 and negative on the nightside). 
As shown in Figure \ref{fig-sw2}, 
\begin{equation}
|h_{\rm eq} - h|_{\rm{day}} + \Delta h + |h_{\rm eq} - h|_{\rm{night}} \sim \Delta h\sub{eq},
\label{eq-daynight}
\end{equation}
where $|h_{\rm eq} - h|_{\rm{day}}$ is the characteristic (scalar) difference between $h\sub{eq}(\lambda,\phi)$ and $h(\lambda,\phi,t)$ on the dayside and $|h_{\rm eq} - h|_{\rm{night}}$ is the characteristic difference on the nightside. Define $|h_{\rm eq} - h|_{\rm{global}}$ to be the arithmetic average between $|h_{\rm eq} - h|_{\rm{day}}$ and $|h_{\rm eq} - h|_{\rm{night}}$.\footnote{The terms $|h_{\rm eq} - h|_{\rm{day}}$ and $|h_{\rm eq} - h|_{\rm{night}}$ are always of the same order because in steady state the rate at which mass is pumped into the active layer on the dayside ($\propto |h_{\rm eq} - h|_{\rm{day}}/\trad$) has to equal the rate at which mass is removed from the nightside ($\propto |h_{\rm eq} - h|_{\rm{night}}/\trad$).} Then, to order of magnitude,
\begin{equation}
|h_{\rm eq}-h|_{\rm{global}} \sim \Delta h_{\rm
   eq} - \Delta h.
\end{equation} 
We thus can approximate $h\sub{eq} -h$ in equation (\ref{mass-cont})
with $\Delta h_{\rm eq} - \Delta h$. The left side of
equation~(\ref{mass-cont}) is, to order of magnitude, $UH/L$, where
$U$ is the characteristic horizontal wind speed and $L$ is the
  characteristic horizontal lengthscale of the flow, which happens to
be of order the planetary radius
$a$.\footnote{\label{footnote-scaling} This scaling is actually
  subtle. Consider the first term on the left-hand side of
  equation~(\ref{mass-cont}).  In a flow where the Rossby number $Ro =
  U/fL \gtrsim 1$, the divergence simply scales as $\nabla\cdot {\bf
    v}\sim U/L$.  In a flow where $Ro\ll 1$, the flow is
  geostrophically balanced; a geostrophically balanced flow has a
  horizontal divergence $-\beta v/f$, and there will be an additional
  possible ageostrophic contribution to the divergence up to order $Ro
  \, U/L$.  Here, $v$ is the meridional (north-south) wind velocity
  and $\beta=df/dy$ is the gradient of the Coriolis parameter with
  northward distance $y$, equal to $2\Omega\cos\phi/a$ on the sphere,
  where $a$ is the planetary radius.  Thus the geostrophic
  contribution to the horizontal divergence is $v\cot\phi/a$.  Because
  the dominant flows on hot Jupiters have horizontal scales comparable
  to the Rossby deformation radius \citep{Showman:2011p12973}, which
  are comparable to the planetary radius for conditions appropriate to
  hot Jupiters, we have that $L\sim a$.  Thus, the first term in
  Equation~(\ref{mass-cont}) scales as $UH/L$ at a typical
  mid-latitude.  Next consider the second term in
  Equation~(\ref{mass-cont}).  When $Ro\gtrsim 1$, this term scales as
  $U\Delta h/L$.  When $Ro\ll1$, geostrophic balance implies that the
  geostrophic component of the flow is perpendicular to $\nabla h$,
  leaving only the ageostrophic component available to flow along
  pressure gradients.  Thus, in this case, the second term scales as
  $Ro \,U\Delta h/L$. Because $\Delta h \lesssim H$, the first term
  generally dominates and the left-hand side of equation~(\ref{mass-cont})
  therefore scales as $UH/L$. } Thus, we have for the continuity
equation
\begin{equation}
H \frac{U}{L} \sim \frac{\Delta h\sub{eq}-\Delta h}{\trad}.
\label{eq_oom1}
\end{equation}

The balance in the momentum equation~(\ref{2DShallowWaterA}) involves
several possibilities. Generally, the
pressure-gradient force $-g \nabla h$,  which drives the flow,
can be balanced by either 
atmospheric drag ($-\mathbf{v}/\tdrag$), the Coriolis force ($-f\mathbf{k} \times \mathbf{v}$), horizontal advection ($\mathbf{v}\cdot\nabla\mathbf{v}$), or the vertical transport term ($\textbf{R}$), which accounts for the momentum transfer from the lower layer. To order of magnitude, the balance is given by
\begin{equation}
\label{eq_oom2}
g\frac{\Delta h}{a} \sim g \frac{\Delta h}{L} \sim \mathrm{max}\left[\frac{U}{\tdrag},fU, \frac{U^2}{L},\frac{U}{H}\frac{\Delta h\sub{eq}-\Delta h}{\trad}\right].
\end{equation}

We solve equations (\ref{eq_oom1}) and
(\ref{eq_oom2}) for the dependent variables $\Delta h$ and $U$.  In the following, we express $\Delta h$ in terms of the dimensionless amplitude $A$ (compare with equation \ref{A_definition}): 
\begin{equation}
A \sim \Delta h/\Delta h\sub{eq}\,.
\end{equation}
We also non-dimensionalize $U$ in terms of the timescale ratio $\tadv/\twave$:
\begin{equation}
\tadv/\twave \sim \sqrt{gH}/U\,.
\end{equation}
There are four possible balances in equation~(\ref{eq_oom2}). Which of the terms is balancing the pressure-gradient force will generally depend on the values of $\trad$, $\tdrag$, the planetary latitude ($\phi$), and the strength of forcing ($\forcing$). Below we solve for the four possible term balances and describe the conditions under which they operate.

\subsection{Drag-dominated: Valid for Both Equatorial and Non-equatorial Regions}

When drag is the dominant term balancing the pressure-gradient
  force, the solutions to equations~(\ref{eq_oom1}) and
(\ref{eq_oom2}) are
\begin{equation}
	A\sim \left(1+\frac{\trad \tdrag}{\twave^2}\right)^{-1},
\label{eq_oom3}
\end{equation}
\begin{equation}
\frac{\tadv}{\twave} \sim \left(\frac{\tdrag}{\twave}\right)^{-1} \left(\frac{\Delta h\sub{eq}}{H}\right)^{-1} \left( 1 + \frac{\trad \tdrag}{\twave^2} \right).
\label{eq_oom3b}	
\end{equation}
Contours of equations (\ref{eq_oom3}) and (\ref{eq_oom3b}) are shown
in the region below the white horizontal line in Figures
\ref{fig-A-model}(c) and (d) ($\tdrag \lesssim \twave$)
and below the white curved line in the right panels of Figure
\ref{fig-equator-forcing}. This white line marks the boundary of the
drag-dominated regime in our model; we will formally define the white
line transition in equation (\ref{eq_white1}).  The same expression
for $A$ as in equation~(\ref{eq_oom3}) results when linearizing the
full model about a state of rest, where $\tadv \rightarrow \infty$
(this linearization is carried out in detail in
\citealt{Showman:2011p12973}).\footnote{To derive equation (\ref{eq_oom3}) from the the shallow water equations, linearize a one-dimensional Cartesian version of equations (\ref{2DShallowWaterA}) and (\ref{2DShallowWaterB}), dropping the Coriolis term. Substitute one equation into the other to eliminate velocity. Impose sinusoidal forcing $h\sub{eq}=\Delta h\sub{eq} \exp(ikx)$ and solve for steady, sinusoidal solutions of the form $h=\Delta h \exp(ikx)$. Finally, solve for $A\sim \Delta h/\Delta h\sub{eq}$ to obtain the same expression as in equation (\ref{eq_oom3}).} This strongly suggests that in this
region of the $\trad,\tdrag$ plane, the advection timescale plays a
minor role in controlling the day-night thickness and temperature
differences. Indeed, Figure \ref{fig-A-model}(d) shows that $\tadv$ is
always significantly larger than $\twave$ in this
region. Equation~(\ref{eq_oom3}) implies that the transition from
small to large $A$ occurs when $\twave \sim \sqrt{\trad\tdrag}$. %
Note that the value of $A$ in equation~(\ref{eq_oom3}) is independent
of the forcing strength $\Delta h\sub{eq}/H$. This is not true for the
characteristic wind speed $U$ and by extension $\tadv$
(equation (\ref{eq_oom3b})).

\subsection{Coriolis-dominated: Valid for Non-equatorial Regions Only}

When atmospheric drag is weak, the dominant balance in the force
equation (\ref{eq_oom2}) depends on the Rossby number, $Ro = U/fL$.
For conditions appropriate to hot Jupiters (typical rotation periods
of a few Earth days, length scales comparable to a planetary radius,
and wind speeds on the order of the wave speed or less), the Rossby
number $Ro \lesssim 1$---except near their equators. For small $Ro$,
the Coriolis force tends to dominate over horizontal advection and 
vertical transport (the last two terms in equation \ref{eq_oom2}).
Thus, away from the equator, we can balance the pressure-gradient
force against the Coriolis force.  In this case, equations
(\ref{eq_oom1}) and (\ref{eq_oom2}) yield

\begin{equation}
	A \sim \left(1+\frac{\trad}{f \twave^2}\right)^{-1},
\label{eq_oom5}
\end{equation}
\begin{equation}
\frac{\tadv}{\twave} \sim \left(\frac{\Delta h\sub{eq}}{H}\right)^{-1} \left(f\twave +\frac{\trad}{\twave}\right).
\label{eq_oom5b}	
\end{equation}
Contours of equations (\ref{eq_oom5}) and (\ref{eq_oom5b}) are shown
in the region above the white line in Figures \ref{fig-A-model}(c) and (d), with the Coriolis parameter $f$ evaluated at a latitude of $\phi=45^{\circ}$. Notice the
similarity between equations (\ref{eq_oom3}) and (\ref{eq_oom5}),
where the role of $\tdrag$ has been replaced by $1/f$. Away from
  the equator, the Coriolis parameter $f\sim \Omega$, and
  equation~(\ref{eq_oom5}) implies that the transition between
  small and large $A$ occurs when $\twave \sim
  \sqrt{\trad/\Omega}$. The transition between the Coriolis-dominated and the
drag-dominated regimes occurs when equations (\ref{eq_oom3}) and (\ref{eq_oom5}) are equal, a condition which yields 
\begin{equation}
\tdrag \sim 1/f.
\label{eq_white1}
\end{equation}
This condition formally defines the white line transition between the Coriolis-dominated and drag-dominated regimes in Figures \ref{fig-A-model}(c) and (d). 

We combine our expressions for $A$ and $\tadv/\twave$ valid away from
the equator (equations~(\ref{eq_oom3}) \& (\ref{eq_oom5}) and (\ref{eq_oom3b}) \& (\ref{eq_oom5b}), together with boundary condition equation~(\ref{eq_white1})) to create Figures \ref{fig-A-model}(c) and (d).
These analytical results (derived for $Ro \lesssim 1$) are a good representation of the globally averaged numerical results shown in Figures \ref{fig-A-model}(a) and (b)---even though the former applies only for non-equatorial regions whereas the latter averages over both non-equatorial and equatorial regions. Nonetheless, the comparison we make between Figures \ref{fig-A-model}(a) and (b) and Figures \ref{fig-A-model}(c) and (d) is fair because either the equatorial solution shows trends in $A$ and $\tadv/\twave$ similar to those of the mid-latitudes, or the equatorial region is small compared to the non-equatorial region (see Section \ref{sec-equatorial}). 

Our analytical theory predicts that $A$ is independent of the forcing strength ($\forcing$) for the entire $\trad,\tdrag$ plane. We test this prediction by running our numerical model at smaller $\forcing$. We show these results in Figure \ref{fig-global-forcing}---indeed all panels exhibit the same general features of Figure \ref{fig-A-model}(a).

\subsection{Advection- or Vertical-transport-dominated: Valid for $Ro\gtrsim1$}
\label{sec-equatorial}

Near the equator (i.e., $Ro \gg 1$), the Coriolis force will vanish 
and the pressure-gradient force will be balanced by either advection or vertical transport. Both possibilities yield the same solution
\begin{eqnarray}\nonumber
A &\sim &1 + \frac{1}{2} \left(\frac{\Delta h\sub{eq}}{H}\right)^{-1} \left(\frac{\trad}{\twave}\right)^2 \times \\ && \left\{ 1 - \left[ 1 + 4 \left(\frac{\Delta h\sub{eq}}{H}\right) \left(\frac{\trad}{\twave}\right)^{-2} \right]^{1/2} \right\},
\label{eq_oom4}
\end{eqnarray}

\begin{eqnarray}\nonumber
&& \frac{\tadv}{\twave} \sim  \\ && \left[ \frac{\Delta h\sub{eq}}{H} + \frac{1}{2} \left(\frac{\trad}{\twave}\right)^2 - \sqrt{\frac{\Delta h\sub{eq}}{H}\left(\frac{\trad}{\twave}\right)^2 + \frac{1}{4} \left(\frac{\trad}{\twave}\right)^4} \right]^{-1/2}.
\label{eq_oom4b}
\end{eqnarray}
We compare these analytical predictions for $A$ at the equator (equations~\ref{eq_oom3} and \ref{eq_oom4}) with the equatorial day-night contrast of our numerical simulations---that is, $A_{\rm{equator}}$---in Figure \ref{fig-equator-forcing} (left panels are the numerical simulations, and right panels are analytical predictions). Notice how the weak drag solution (\ref{eq_oom4}) depends on forcing amplitude $\forcing$, whereas the strong drag solution (\ref{eq_oom3}) does not. The white line---marking the transition between the weak- and strong-drag regimes---is now a parabola, obtained by equating (\ref{eq_oom3}) and (\ref{eq_oom4}). As $\forcing$ is reduced, wind speeds are reduced and a greater region of phase space is drag dominated. 

For the strong forcing expected on hot Jupiters ($\forcing \sim 1$),
the numerical solution for the day-night contrast at the equator $A_{\rm{equator}}$ (lowermost panels in Figure \ref{fig-equator-forcing}) is very similar to the one obtained for mid-latitudes (Figures \ref{fig-A-model}(a) and (c)), deviating by at most $\sim$$15 \%$. 
For weaker forcing, the solutions valid at mid-latitudes and the
equator differ; however, at weak forcing, equations
(\ref{eq_oom4}) and (\ref{eq_oom4b}) are only valid in a narrow
range of latitudes centered on the equator. This latitudinal range is delimited by the condition $Ro = 1$ and can be found analytically by solving 
\begin{eqnarray}\nonumber
&& Ro=\frac{U}{fL}\sim 1\sim  \left(\frac{\sqrt{gH}}{2\Omega a}\right)^{1/2} \frac{1}{\sin\phi} \times\\
&& \left[ \frac{\Delta h\sub{eq}}{H} + \frac{1}{2} \left(\frac{\trad}{\twave}\right)^2 - \sqrt{\frac{\Delta h\sub{eq}}{H}\left(\frac{\trad}{\twave}\right)^2 + \frac{1}{4} \left(\frac{\trad}{\twave}\right)^4} \right]^{1/2}
\label{eq_Rossby}
\end{eqnarray}
for $\phi$. For $\forcing = (1,0.1,0.01,0.001)$ the width of the equatorial region is at most $\phi \sim (\pm 30^{\circ},\pm 10^{\circ}, \pm 3^{\circ}, \pm 1^{\circ})$ and goes to zero in the limit that the forcing amplitude
goes to zero.

\section{INTERPRETATION OF THEORY}
\label{sec-discussion}

\subsection{Timescale Comparison}
\label{sec-timescale}

\begin{figure}[ht]
\centering
\includegraphics[width=\linewidth]{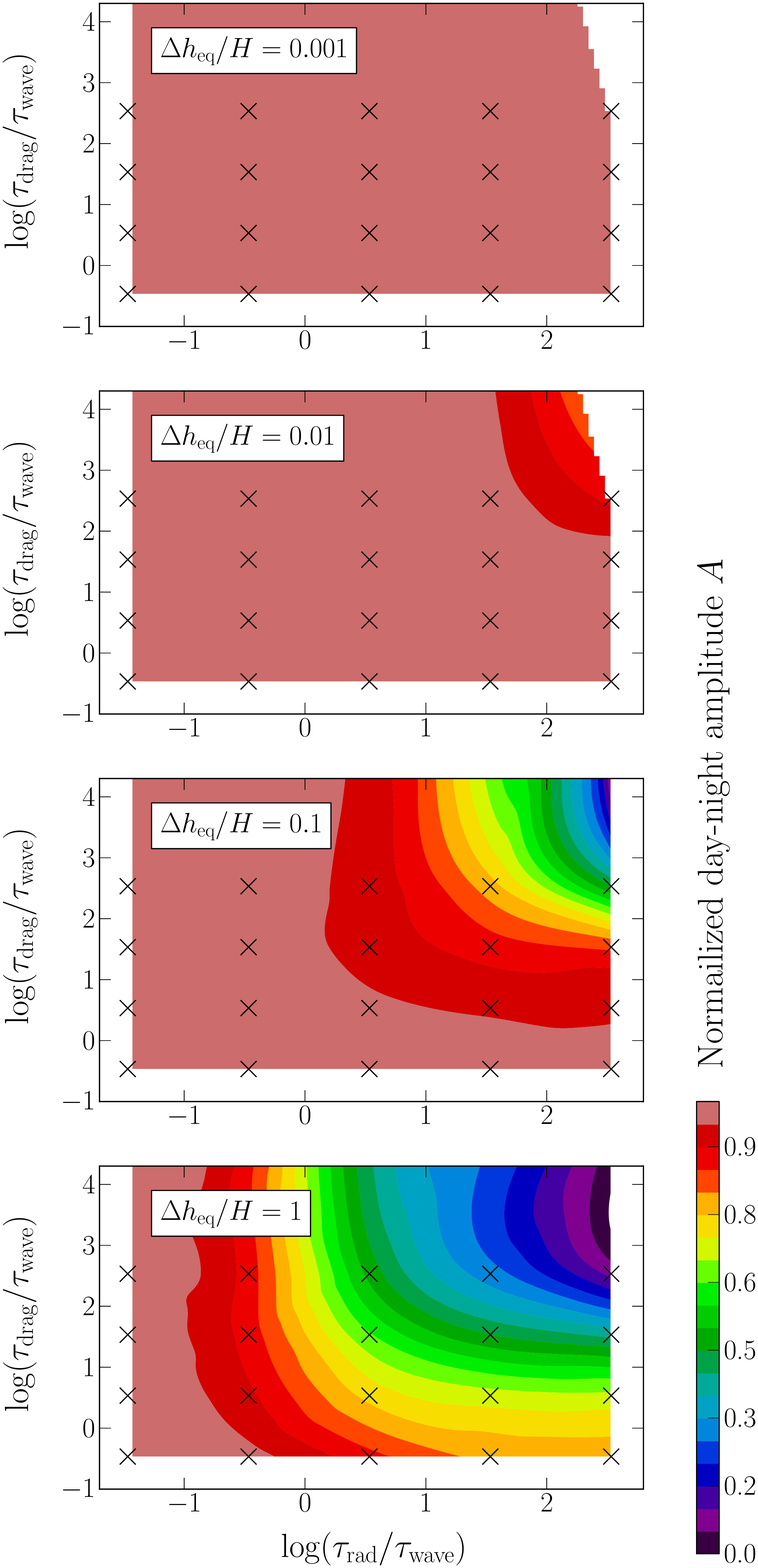}
\caption{%
Contours of day-night height amplitude $A$ as would be predicted by comparing $\trad$ vs. $\tadv$---the horizontal advective timescale. Values for $A$ were computed with equation (\ref{eq-A-trad-tadv}), where $\tadv \equiv L/U$, and $U$ set to the global RMS value of the wind speed in the shallow water model. At $\forcing=1$ (lowermost panel), the advective term in the momentum equation is of the same order of magnitude as other terms. As a result, comparison between $\tadv$ and $\trad$ yields $A$-contours that show some similarity to the numerical results (Figure \ref{fig-A-model}(a)). But as $\forcing$ is reduced, the advective term dies off faster than other terms, making advection less relevant (with the exception of a thin band around the equator). By the time $\forcing =0.001$, $\tadv$ has become so large compared to $\trad$ that equation (\ref{eq-A-trad-tadv}) predicts $A \to 1$ over the entire parameter space, clearly contradicting numerical results. The low forcing amplitude cases demonstrate that the height field in the shallow water model is not being predominantly redistributed by planetary-scale horizontal advection.}
\label{fig-trad-vs-tadv}
\end{figure}

To give an executive summary of Section \ref{sec-theory}: we can reproduce the characteristic day-night difference $A(\trad,\tdrag,\forcing)$ with a set of simple scaling relations 
\begin{equation}
A \sim \left\{
  \begin{array}{l l}
    \left(1+\frac{\trad \tdrag}{\twave^2}\right)^{-1} & \quad \text{when $\tdrag \lesssim \Omega^{-1}$}\\
    \left(1+\frac{\trad}{\Omega \twave^2}\right)^{-1}& \quad \text{when $\tdrag \gtrsim \Omega^{-1}$,}\\
  \end{array} \right.
\label{eq-A-combined}
\end{equation}
which are valid for all forcing strengths ($\forcing$) and nearly all latitudes (except those closest to the equator where $A$ is given by equation (\ref{eq_oom4}) in the weak drag limit).
We have found both numerically and analytically that a transition from a planet with uniform atmospheric temperature ($A\sim 0$) to one with a large day-night temperature contrast relative to radiative equilibrium ($A\sim 1$) occurs when
\begin{equation}
 \left\{
  \begin{array}{l l}
    \twave \sim \sqrt{\trad \tdrag} & \quad \text{when $\tdrag \lesssim \Omega^{-1}$}\\
    \twave \sim \sqrt{\trad/\Omega} & \quad \text{when $\tdrag \gtrsim \Omega^{-1}$.}\\
  \end{array} \right.
\label{eq-A-trans}
\end{equation}
Wave adjustment would thus seem to play a key role in controlling
whether or not the thermal structure of the day-night contrast is
close to radiative equilibrium. In contrast, horizontal advection and
radiative damping are usually considered the dominant factors for heat
redistribution. The comparison between $\tadv$ and $\trad$ is
intuitive and provides a reasonable estimate for the heat
redistribution efficiency on hot Jupiters (e.g.,
\citealt{Perna:2012p17536}), which are strongly forced ($\forcing \sim
1$). Nevertheless, the $\tadv$ vs. $\trad$ comparison is a poor
predictor for $A$ in the more general case, where $\forcing$ is not of
order unity. We now show this explicitly.

In Figure \ref{fig-trad-vs-tadv} we show $A$-contours as would be predicted by the $\trad$ vs. $\tadv$ timescale comparison: 
\begin{equation}
	A\sim  \left(1+ \frac{\trad}{\tadv}\right)^{-1}.
\label{eq-A-trad-tadv}
\end{equation}
We chose the functional form for $A$ in equation
(\ref{eq-A-trad-tadv}) because it possesses the correct limiting
values (including $A=1/2$ when $\trad=\tadv$) and allows for a direct
comparison with our results.  Note that $\tadv \propto U^{-1}$ is not
an input parameter in the model; it has to be estimated either from
the numerical solution or by using our scaling solutions (equations
(\ref{eq_oom3b}), (\ref{eq_oom5b}), or (\ref{eq_oom4b})). In Figure
\ref{fig-trad-vs-tadv}, we set $U$ equal to the global RMS value of
the wind speed in the numerical model. At large forcing amplitude
($\forcing =1$, lowermost panel) the contours of $A$ as predicted by
equation (\ref{eq-A-trad-tadv}) show some resemblance to the numerical
shallow water solution shown in Figure \ref{fig-global-forcing}
(lowermost panel). Nevertheless, as the forcing amplitude is reduced,
the values of $A$ predicted by equation (\ref{eq-A-trad-tadv}) become
increasingly inaccurate. In the limit where $\forcing \rightarrow 0$, equation
(\ref{eq-A-trad-tadv}) predicts $A \rightarrow 1$ everywhere, because
the characteristic wind speed $U\propto \forcing$ (see equations
(\ref{eq_oom3b}) and (\ref{eq_oom5b})). In contrast, our numerical results
show that $A$-contours are largely independent of $\forcing$.

Rather than invoking Equation~(\ref{eq-A-trad-tadv}), we can
  demonstrate the breakdown of the $\trad$-vs-$\tadv$ prediction in
  the low-amplitude limit simply by considering the ratio of these two
  timescales.  When $\forcing=0.001$, values of $\tadv/\trad$ vary
  between $30$ and $3 \times 10^5$ over our explored parameter space.
  Therefore, a $\trad$-vs-$\tadv$ comparison would predict that the
  day-night thickness contrast is always very close to
  radiative equilibrium over the entire explored parameter space.
  This is inconsistent with our numerical simulations
  (Figures~\ref{fig-fullsolution-0p001} and \ref{fig-global-forcing}, 
  top panel), which clearly show a transition from models close
  to radiative equilibrium at short $\trad$ to models with much smaller
  thermal contrasts at long $\trad$.

It is clear that, at least for low forcing amplitudes, the amplitude
of the day-night thermal contrast (relative to radiative equilibrium)
is not controlled by a comparison between $\tau_{\rm rad}$ and
$\tau_{\rm adv}$. We now give two physical interpretations of the
theory.

\subsection{Vertical Advection}

The timescale comparison in equation (\ref{eq-A-trans}) can be
obtained by comparing a \textit{vertical} advection timescale to the
radiative timescale. Define the vertical advection time, $\tvert$, as
the time for a fluid parcel to move vertically over a distance
corresponding to the day-night thickness difference $\Delta h$. The
vertical velocity, by mass continuity, is $\sim$$H \, \nabla \cdot
\textbf{v}$ (where $\nabla$ is the horizontal gradient operator and
$\textbf{v}$ is the horizontal velocity). Then
\begin{equation}
\tvert \sim \frac{\Delta h}{H \, \nabla \cdot \textbf{v}} \sim \frac{\Delta h L}{HU}.
\label{eq-tvert1}
\end{equation}

In the strong drag regime, equation (\ref{eq_oom2}) becomes $\Delta h /U \sim L/(g\tdrag)$, which when substituted into equation (\ref{eq-tvert1}) implies that
\begin{equation}
\tvert \sim  \frac{L^2}{gH \tdrag} \sim \frac{\twave^2}{\tdrag}.
\label{eq-tvert2}
\end{equation}
If $\trad$ is of order $\tvert$, then
\begin{equation}
\trad \sim \tvert \sim \frac{\twave^2}{\tdrag}, 
\label{eq-tvert2b}
\end{equation}
which is precisely the same comparison in equation (\ref{eq-A-trans})
derived in the strong drag limit. Thus, our solution approaches
radiative equilibrium ($\Delta h \rightarrow \Delta h\sub{eq}$) when
$\tvert \gg \trad$, i.e., when $\twave \gg \sqrt{\trad \tdrag}$. In
other words, the atmosphere is close to radiative equilibrium when the
vertical advection time (over a distance $\Delta h$) is long compared
to the radiative time. Conversely, the behavior is in the
limit of small thickness variations ($\Delta h \ll \Delta h_{\rm eq}$)
when $\tvert \ll \trad$, i.e., when $\twave^2 \ll \trad\tdrag$.
In other words, the day-night thickness difference is small (compared to
radiative equilibrium) when the vertical advection time is short compared
to the radiative time.

In the Coriolis-dominated regime, the role of $\tdrag$ is replaced by $f^{-1} \sim \Omega^{-1}$, as can readily be seen in equation (\ref{eq_oom2}). Setting $\trad$ equal to $\tvert$ now yields  
\begin{equation}
\trad \sim \tvert \sim \Omega \twave^2, 
\label{eq-tvert3}
\end{equation}
which is the same timescale comparison in equation (\ref{eq-A-trans})
for the Coriolis-dominated regime.

In either case, the relationship between the vertical and
  horizontal advection times follows directly from
  equation~(\ref{eq-tvert1}):\footnote{This relationship can also
be obtained by equating the rate of vertical mass
transport $\sim$$\rho\sub{upper} \Delta h L^2/\tvert$ to the rate
of horizontal mass transport $\sim$$\rho\sub{upper} H L^2/\tadv$.}
\begin{equation}
	\tvert \sim \tadv \frac{\Delta h}{H}.
\label{eq-tvert1b}
\end{equation}
Therefore, the vertical advection time, $\tvert$, is smaller than the horizontal advection time by $\Delta h/H$.
The vertical and horizontal advection times become comparable when day-night thickness differences are on the order of unity (which can occur only for large forcing amplitudes $\forcing \sim 1$). Only in that special case, will a comparison between $\trad$ and the horizontal advection time, $\tadv$, give a roughly correct prediction for $A$ (as previously mentioned in Section \ref{sec-timescale} and shown in the bottom panel of Figure \ref{fig-trad-vs-tadv}).

The physical reason for the importance of $\tvert$ over
  $\tadv$ in controlling the transition stems from the relative roles
  of vertical and horizontal advection in the continuity equation
  (\ref{mass-cont}).  When $\Delta h/H$ is small, the term
  $h\nabla\cdot {\bf v}$---which essentially represents vertical
  advection---dominates over the horizontal advection term ${\bf
    v}\cdot \nabla h$ (see footnote~\ref{footnote-scaling}).  Thus,
  under conditions of small $\Delta h/H$, the dominant balance is
  between radiative heating/cooling and {\it vertical}, rather than
  horizontal, advection.  This is precisely the shallow-water version
  of the so-called ``weak temperature gradient'' (WTG) regime that
  dominates in the Earth's tropics, where the time-mean balance in the thermodynamic
  energy equation is between vertical advection and radiative cooling
\citep{Sobel:2001p14821, Bretherton:2003p15030}.  Only when
$\Delta h/H\sim 1$ does horizontal advection generally become
comparable to vertical advection in the local balance.

\subsection{Wave Adjustment Mechanism}

As described in Section~\ref{sec-intro}, gravity waves are known to
play a central role in regulating the thermal structure in the Earth's
tropics \citep{Bretherton:1989p15029, Sobel:2002p17814,
  Showman:2013p17094}. Likewise, our analytic scaling theory
(Section~\ref{sec-theory}) shows the emergence of a wave timescale in
controlling the transition from small to large day-night temperature
difference in our global, steady-state solutions.  This is not
accidental but strongly implies a role for wave-like processes in
governing the dynamical behavior.  Although our numerical simulations and theory
stand on their own, we show here that they can be interpreted in
terms of a wave adjustment mechanism.

To illustrate, consider a freely propagating gravity wave in the
time-dependent shallow-water system.  In such a wave, horizontal
variations in the thickness ($h$) cause pressure-gradient forces that
induce horizontal convergence or divergence, which locally changes the
height field and allows the wave structure to propagate laterally.  In
fact, it can easily be shown that this physical process---namely,
vertical stretching/contraction of atmospheric columns in response to
horizontal pressure-gradient forces---naturally leads to a wave
timescale $\tau_{\rm wave}\sim L/\sqrt{gH}$ for a freely propagating
wave to propagate
over a distance $L$.\footnote{Specifically, consider for simplicity a
  one-dimensional, non-rotating linear shallow-water system in the
  absence of forcing or damping.  If a local region begins with a
  height $\Delta h$ different from surrounding regions, the timescale
  for the wave to propagate over its wavelength $L$ will be determined
  by the time needed for horizontal convergence/divergence to locally
  change the height by $\Delta h$.  This timescale is simply given by
\begin{equation}
\tau \sim {\Delta h\over H\nabla\cdot{\bf v}}\sim {\Delta h \, L\over H U}\, ,
\end{equation}
where $U$ is the wind speed associated with the wave motion.  In
the absence of forcing or damping, the horizontal momentum equation
is the linearized version of Equation~(\ref{2DShallowWaterA}) with the right-hand
side set to zero.  In the absence of rotational effects, the
balance is simply between the time-derivative term and the pressure-gradient
term, which to order of magnitude is
\begin{equation}
{U\over \tau} \sim g{\Delta h\over L}\, .
\end{equation}
Combining these two equations immediately yields $\tau\sim {L\over
  \sqrt{gH}}$.}  Now, although our solutions in
Section~\ref{sec-numerical-solutions} lack freely propagating waves
(being forced, damped, and steady), the key point is that the {\it
  same physical mechanism} that causes wave propagation in the free,
time-dependent case regulates the day-night thickness
variation in our steady, forced, damped models.  In particular, in
our hot-Jupiter models, the
dayside mass source and nightside mass sink cause a thickening of the
layer on the dayside and a thinning of it on the nightside; in
response, the horizontal pressure-gradient forces cause a horizontal
divergence on the dayside and convergence on the nightside, which
attempts to thin the layer on the dayside and thicken it on the
nightside.  Although technically no phase propagation occurs in this
steady case, this process---being essentially the same process that
governs gravity-wave dynamics---nevertheless occurs on the gravity-wave
timescale.  This is the reason for the emergence of $\tau_{\rm wave}$
in our analytic solutions in Section~\ref{sec-theory}.

The situation on a rotating planet is more complex, because
the Coriolis force significantly modifies the wave behavior.
On a rotating planet, freely propagating, global-scale waves within a
deformation radius of the equator split into a variety of equatorially
trapped wave modes, including the Kelvin wave and equatorially trapped
Rossby waves. (For overviews, see \citet{Matsuno:1966p0001}, \citeauthor{holton2004introduction} (2004,
  pp.~394--400) or \citeauthor{andrews1987middle} (1987, pp.~200--208).)  The Kelvin
wave exhibits pressure perturbations peaking at the equator, with
strong zonally divergent east-west winds; these waves propagate to the
east.  The east-west winds in the Kelvin wave cause strong north-south
Coriolis forces that prevent the expansion of the pressure perturbations
in latitude; however, nothing resists the pressure perturbations
in longitude, and so the Kelvin wave propagates zonally like a gravity
wave at a speed $\sqrt{gH}$.
In contrast, the Rossby wave exhibits pressure perturbations that peak
off the equator and vortical winds that encircle these pressure
perturbations; these waves propagate to the west. See \citeauthor{holton2004introduction} (2004, Figure
11.15) and \citeauthor{Matsuno:1966p0001} (1966, Figure 4(c)), respectively, for visuals of
these two wave types.  

The link between our solutions and wave dynamics becomes even tighter
when one compares the detailed spatial structure of the solutions to
these tropical wave modes.  Building on a long history of work in
tropical dynamics (e.g., \citealt{Matsuno:1966p0001, Gill:1980p12543}), \citet{Showman:2011p12973} showed that the behavior of steady, forced, damped solutions
like those in
Figures~\ref{fig-fullsolution} and \ref{fig-fullsolution-0p001} can be
interpreted in terms of standing, planetary-scale Rossby and Kelvin
waves. Examining, for example, the $\tau_{\rm rad}=1\rm\,day$,
$\tau_{\rm drag}=\rm 1\,day$ models in
Figure~\ref{fig-fullsolution-0p001}, the off-equatorial behavior,
including the off-equatorial pressure maxima and the vortical winds
that encircle them (clockwise in northern hemisphere, counterclockwise
in southern hemisphere), is dynamically analogous to the equatorially
trapped Rossby wave mentioned above.  The low-latitude behavior, with
winds that zonally diverge from a point east of the substellar point,
is dynamically analogous to the equatorial Kelvin wave mentioned
above. See \citet{Gill:1980p12543} and \citet{Showman:2011p12973}
for further discussion.
Again, here is the key point: the physical mechanisms that
cause stretching/contraction of the shallow-water column and wave
propagation of Kelvin and Rossby waves in the free, time-dependent
case are {\it the same physical mechanisms} that regulate the spatial
variations of the thickness in our forced, damped, steady solutions.

In sum, if the radiative and frictional damping times are sufficiently 
long, the Kelvin and Rossby waves act efficiently to flatten the layer, and the
day-night thickness differences are small.  If the radiative and
frictional damping times are sufficiently short, the Kelvin and Rossby
waves are damped and cannot propagate zonally; the thermal structure
is then close to radiative equilibrium.  Although our solutions are
steady, this similarity to wave dynamics explains the natural
emergence of a wave timescale in controlling the transition
between small and large day-night contrast.  In particular, because
the Kelvin wave propagates at a speed $\sqrt{gH}$, the fundamental
wave timescale that emerges is  $L/\sqrt{gH}$.

It is important to emphasize that, despite the importance of wave timescales, 
horizontal advection nevertheless plays
a crucial role in the dynamics.  Consider an imaginary surface at the terminator
dividing the planet into dayside and nightside hemispheres.  It is horizontal
advection across this surface that ultimately transports heat from day to night, thereby
allowing each hemisphere to reach a steady state in the presence of continual
dayside heating and nightside cooling.  In the linear limit of our shallow-water
model, this transport manifests as advection of the mean thickness (i.e., $uH$ integrated
around the terminator), although advection of thickness {\it variations} can also
play a role at high amplitude, when these variations are not small relative to $H$.
The importance of horizontal advection does not mean that the flow behavior
is controlled by the horizontal advection timescale, and indeed we have shown
that it is generally not, particularly when the forcing amplitude is weak.

\section{APPLICATION TO HOT JUPITER OBSERVATIONS} 
\label{sec-observations}

Here we compare predictions for the day-night height difference, $A$, obtained from our shallow water model to the observed fractional infrared flux variations, $A\sub{obs}$, on hot Jupiters. Because the shallow water equations do not explicitly include stellar irradiation, we have to express our model's input parameters $\trad$ and $\tdrag$ in terms of $T\sub{eq}$ (our proxy for stellar irradiation). We find the dependence of the radiative timescale on $T\sub{eq}$ by approximating
$\trad$ as the ratio between the available thermal energy per
unit area within a pressure scale-height and the net radiative
flux from that layer \citep{Showman:2002p12764},

\begin{equation}
\trad \sim \frac{P c\sub{P}}{4 g \sigma T\sub{eq}^3}, 
\label{trad_estimate}
\end{equation}

\noindent where $P$ is the atmospheric pressure at the emitting layer, $c\sub{P}$ is the specific heat, and $\sigma$ is the Stefan-Boltzmann constant.
We have chosen to leave $\tdrag$ as a free parameter, because the source of atmospheric drag in gas giants remains largely unknown (e.g., \citealt{Perna:2010p14307, Li:2010p14315, 2010exop.book..471S}), and because our results suggest a weak dependence of $A$ on $\tdrag$ (see Figure \ref{fig-A-model}(a)).\footnote{In the case of hot Jupiters, the main candidate for atmospheric drag is Lorentz-force breaking of the thermally ionized atmosphere. For the case of HD 209458b, \citet{Perna:2010p14307} estimate that $\tdrag$ could reach values as low as $\sim$$0.1$ days on the planet's dayside, which matches our lowest considered $\tdrag$. \citet{Rauscher:2013p17720} found that the inclusion of Lorentz drag in their global circulation model of HD 209458b changes the ratio between maximum and minimum flux emission from the planet by up to 5\% when compared to a drag-free model (see their Table 2). \revd{Although their study excluded the most strongly irradiated hot Jupiters,} their simulations are in accordance with our result that $A$ depends only weakly on $\tdrag$.}

\begin{figure}
\includegraphics[width=\linewidth]{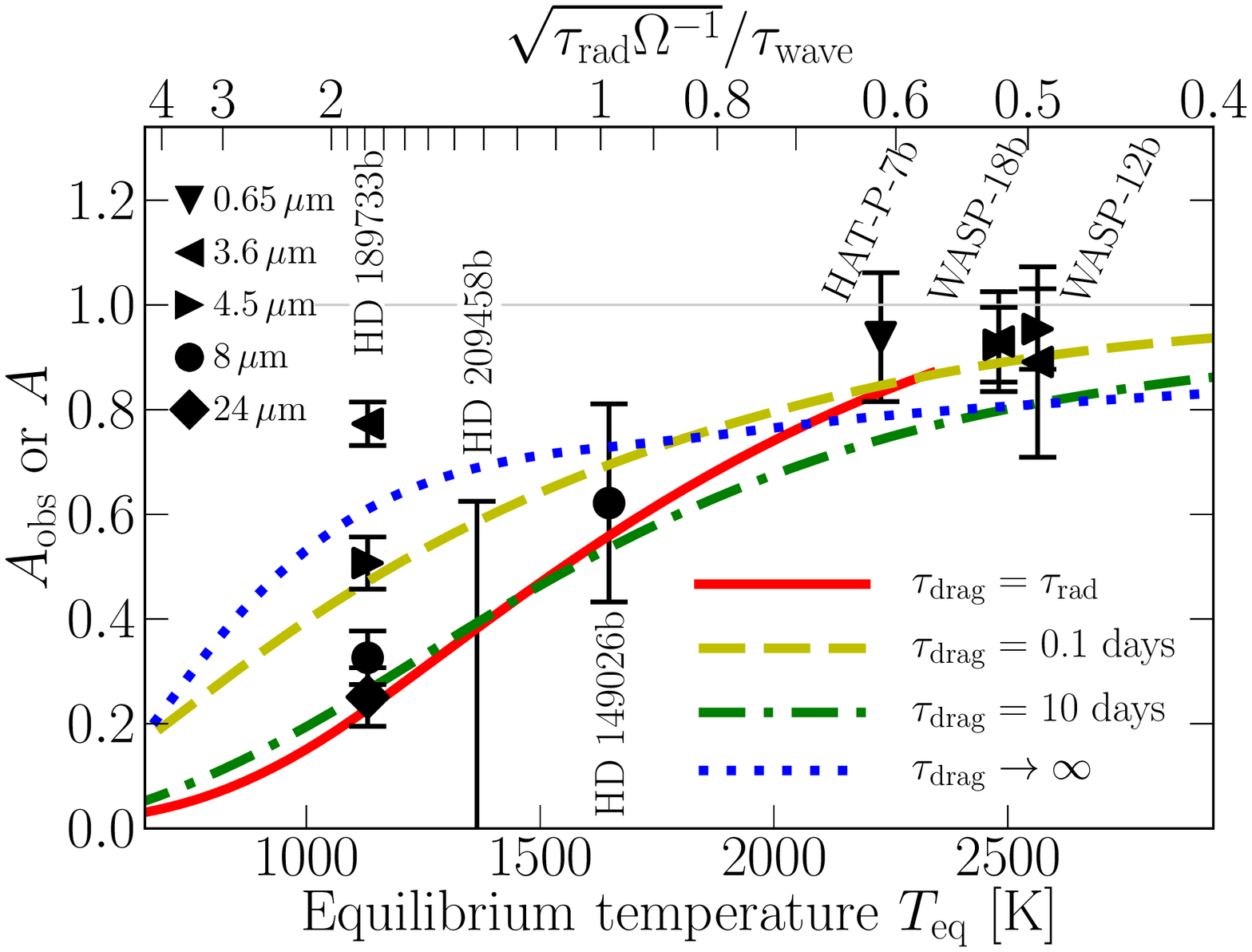}
\caption{%
Same as Figure \ref{fig-obs-nomodel}, but including shallow water model results. Black symbols are fractional day-night flux
variations ($A\sub{obs}$) for hot Jupiters with measured light curves, as explained in detail in  Figure \ref{fig-obs-nomodel}. Colored dashed curves show model results for the
normalized day-night height difference $A$
  for constant $\tau\sub{drag}$, expressed in Earth days, while the
  solid red line shows the solution when
  $\tau\sub{drag}=\tau\sub{rad}$. Equilibrium temperatures ($T\sub{eq}$) for the
  model were estimated with equation~(\ref{trad_estimate}) with $c\sub{P}=10^4$ J~kg$^{-1}$~K$^{-1}$, $g=10$ m~s$^{-2}$, and $P=0.25$ bar, which is approximately the pressure of the layer radiating to space (compare upper and lower $x$-axes). Because plotted solutions are (mostly) in the weak-drag regime, all curves lie close together and broadly reproduce the observational trend of increasing $A\sub{obs}$ with increasing $T\sub{eq}$.}
\label{fig-obs-model}
\end{figure}

We plot model values for $A$ together with $A\sub{obs}$ from hot Jupiter observations in Figure \ref{fig-obs-model}. Each broken curve shows model results for $A$ for a constant value of $\tdrag$, ranging from values that are marginally in the strong drag regime ($\tdrag = 0.1$ days) to no drag ($\tdrag \to \infty$). Additionally, we show the solution when $\tdrag=\trad$ with the solid red line. 
Because $\tdrag$ is fixed, the only remaining free variable is $\trad$, which we express in terms of $T\sub{eq}$ using equation (\ref{trad_estimate}) (compare upper and lower $x$-axes).
All curves follow roughly the same path which reproduces the observational trend of increasing $A$ with equilibrium temperature $T\sub{eq}$. The curves run close to each other because the models considered are either in the weak drag regime or barely in the strong drag regime. \revd{Solutions with $\tdrag < 0.1$ days are numerically challenging and were therefore not explored. In any case, for the case of magnetic drag, the temperatures required to reach such low drag timescales are $\gtrsim$1500 K \citep{Perna:2010p14307}, where the corresponding $\trad$ is already so low that $A\sim 1$ regardless of the strength of drag.} 
\revd{In summary,} when $\sqrt{\trad/\Omega}$ is shorter than $\twave$, the Kelvin and Rossby waves that emerge near the equator are damped before they can propagate zonally, resulting in $A \sim 1$. In contrast, when $\sqrt{\trad/\Omega}$ is long compared to $\twave$, the emerging waves can propagate far enough to flatten the fluid layer, resulting in $A \sim 0$.

\section{CONCLUSIONS}
\label{sec-conclusions}

We have presented a simple atmospheric model for tidally locked exoplanets that reproduces the observed transition from atmospheres with longitudinally uniform temperatures to atmospheres with large day-night temperature gradients as stellar insolation increases (Figure \ref{fig-obs-model}). 
In our model we have parameterized the stellar insolation in terms of a radiative timescale, $\trad$, and frictional processes in terms of a drag timescale, $\tdrag$. The shallow water model contains two additional natural timescales: the rotation period of the planet ($\sim$$\Omega^{-1}$), and $\twave$, the timescale over which gravity waves travel horizontally over planetary distances. 
We have developed an analytical scaling theory to estimate the heat redistribution efficiency in terms of these four timescales. %
Our scaling theory predicts that for sufficiently weak atmospheric
drag, the temperature distribution on the planet can be estimated by
the ratio of $\twave$ and $\sqrt{\trad/ \Omega}$. Drag will influence the day-night
temperature contrast if it operates on a timescale shorter than
$\Omega^{-1}$. In this drag-dominated regime, the heat redistribution efficiency will depend on the ratio of $\twave$ and $\sqrt{\trad \tdrag}$. These scaling relations 
are summarized in equation (\ref{eq-A-trans}). We provide two physical interpretations to understand why these timescales arise from the shallow water model. 
We derive the first interpretation by noting that the same physical mechanisms that generate equatorially trapped waves in an undamped shallow water model also regulate the steady-state solutions of our forced-damped model. The heat redistribution efficiency is therefore related to the characteristic distance that waves can travel before they are damped. This distance is set by the relative magnitudes of the timescale for waves to travel over planetary distances, $\twave$, and the timescale for the waves to damp.  

For the second interpretation, we recognize that the timescale
comparisons of equation (\ref{eq-A-trans}) can be written as $\trad
\sim \tvert$, where $\tvert$ is the timescale for a parcel to advect
vertically over a distance equal to the day-night difference in
thickness. This criterion emerges from the fact that, as long as
  forcing amplitudes are not large, it is primarily vertical
advection---and not horizontal advection, as commonly
assumed---that balances radiative relaxation in the continuity equation.
Despite this fact, the timescale comparison between $\trad$ and the
horizontal advection timescale, $\tadv$, provides reasonable estimates
for the heat redistribution efficiency on hot Jupiters. This is
because these gas giants have strongly forced atmospheres ($\forcing
\sim 1$), where $\tadv$ becomes comparable to $\tvert$ (see equation
(\ref{eq-tvert1b})), and where horizontal and vertical thermal
  advections become comparable. In weakly forced systems, the
timescale comparison between $\trad$ and $\tadv$ is a poor predictor
for the heat redistribution efficiency, as we show in Figure
\ref{fig-trad-vs-tadv}. In contrast, the timescale comparison between
$\trad$ vs. $\tvert$---derived from the dynamical equations---yields a
more accurate estimate of the heat redistribution efficiency at any
$\forcing$, including the strongly forced hot Jupiters.

\begin{acknowledgments}
We are extremely thankful to Eugene Chiang, whose extensive comments and manuscript revisions greatly helped to clarify the presentation of this work. This project was initiated at the International Summer Institute for Modeling in Astrophysics (ISIMA) at the Kavli Institute
for Astronomy and Astrophysics (KIAA) at Peking University.  We
thank Pascale Garaud and Doug Lin for their help in organizing
the program. \revd{We are also grateful to an anonymous referee for promoting a more thorough discussion on the limitations of our model.} This work was supported by NASA Origins grants
NNX08AF27G and NNX12AI79G to A.P.S., and by graduate fellowships from the National Science Foundation and UC MEXUS/CONACyT awarded to D.P.-B.

\end{acknowledgments}

\def\icarus{Icarus}
\def\nat{Nature}
\def\apj{Astrophys. J.}
\def\mnras{Mon. Not. Roy. Astron. Soc.}
\def\aap{Astron. Astrophys.}
\def\grl{Geophys. Res. Lett.}
\def\jgr{J. Geophys. Res.}
\def\upsi{$\upsilon$}
\bibliographystyle{apj} 
\bibliography{adsdpb}

\end{document}